\documentclass[%
 reprint,
nofootinbib,
 amsmath,amssymb,
 aps,
]{revtex4-1}

\usepackage{graphicx}
\usepackage{dcolumn}
\usepackage{braket}
\usepackage{gensymb}
\usepackage{mathtools}
\usepackage{bm}
\usepackage{amsmath}
\usepackage{float}
\usepackage{color}
\usepackage{units}
\usepackage{lipsum}
\usepackage{csquotes}
\usepackage[makeroom]{cancel}
\usepackage{multirow}
\usepackage{booktabs}
\usepackage{amssymb}
\usepackage{color}

\newcommand{\drma}[1]{{\color{black} #1}}
\newcommand{\last}[1]{{\color{black} #1}}

\makeatletter
\newcommand*{\rom}[1]{\expandafter\@slowromancap\romannumeral #1@}
\makeatother

\DeclarePairedDelimiter\floor{\lfloor}{\rfloor}

\usepackage[utf8]{inputenc}

\begin{document}

\preprint{APS/123-QED}

\title{Evaluation of Counterfactuality in Counterfactual Communication Protocols}
\author{D. R. M. Arvidsson-Shukur}
\affiliation{%
 Cavendish Laboratory, Department of Physics, University of Cambridge, CB3 0HE, Cambridge, United Kingdom
}%

\author{C. H. W. Barnes}%
\affiliation{%
 Cavendish Laboratory, Department of Physics, University of Cambridge, CB3 0HE, Cambridge, United Kingdom
}%

\author{A. N. O. Gottfries}
\affiliation{%
 Faculty of Economics, University of Cambridge, CB3 9DD, Cambridge, United Kingdom
}%

\date{\today}

\begin{abstract}
We provide an in-depth investigation of parameter estimation in Nested Mach-Zehnder interferometers (NMZIs) using two information measures: the Fisher information and the Shannon mutual information. Protocols for counterfactual communition (CFC) have, so far, been based on two different definitions of counterfactuality. In particular, some schemes are been based on NMZI devices, \drma{and} have recently been subject to criticism. We provide a methodology for evaluating the counterfactuality of these protocols, based on an information theoretical framework. More specifically, we make the assumption that any realistic quantum channel in MZI structures will have some weak uncontrolled interaction. We then use the Fisher information \drma{of this interaction} to measure counterfactual violations. The measure is used to evaluate the suggested counterfactual communication protocol of Salih et al. \cite{Salih13}. The protocol of Arvidsson-Shukur and Barnes \cite{ArvShukur16}, based on a different definition, is evaluated with a probability measure. Our results show that the definition of Arvidsson-Shukur and Barnes is satisfied by their scheme, whilst that of Salih et al. is only satisfied by perfect quantum channels. For realistic devices the latter protocol does not achieve its objective.
\begin{description}
\item[PACS numbers]
03.65.Ta, 03.67.Hk, 03.67.Ac
\end{description}
\end{abstract}

\pacs{Valid PACS appear here}
\maketitle


\section{Introduction}

During the past one and a half centuries, the study of interferometers has resulted in some of the most profound discoveries in physics. From the Michelson–Morley experiment \cite{Michelson1887}, which established the speed of light as a constant, to Hardy's Paradox \cite{Hardy92}, which elegantly demonstrates the non-local behaviour of the fundamentals of quantum physics, interferometers have played a pivotal role. This is perhaps more evident today than ever before, considering the recent discoveries of gravitational waves made with two power-recycled Michelson interferometers.\cite{Abbott16}

Studies with optical quantum interferometers have shown great promise, not only for the detection of novel physics, but also for external field detection and external parameter estimation.\cite{Hariharan03, Bertocchi06, Berti07, Stockton07, Dorner09, Cable10, Vitale10, Bahder11, Rozema12, Rozema12-2, Mahler13} A common example of this is phase estimation. By letting optical quantum states interfere with a medium inside an interferometer it can be easier to establish the nature of a phase-shift caused by the medium, than with a \textit{direct} interaction.\cite{Hradil95, Bahder06, Simon08, Dorner09} 

A framework for studying phase estimation in interferometers has been developed by Bahder et al. \cite{Bahder06, Simon08, Bahder11}. This framework uses the Shannon mutual information and the classical Fisher information as measures of the phase estimating capacity of different interferometers and input states. The Shannon mutual information provides a measure of the suitability of a specific experiment for the estimation of a phase given some known or unknown phase probability distribution. The classical Fisher information, on the other hand, provides a measure of how well a specific---but unknown---phase-shift in the interferometer can be estimated from the outcome events of the specific interferometry experiment.

Another area of physics that has been developed entirely via the study of interferometers, is that of counterfactual phenomena:

\begin{displayquote}
\textit{\drma{C}ounterfactuals} -- that is, things that might have happened, although they did not in fact happen\drma{.}

 \drma{--} Roger Penrose in `Shadows of the Mind' \cite{bPenrose94}
\end{displayquote}

 The counterfactual phenomenon of interaction-free measurements was originally discovered by Elitzur and Vaidman in their seminal paper on quantum bomb diffusal.\cite{Elitzur93} They showed how a Mach-Zehnder interferometer (MZI) could be used in order to query whether or not an absorbing object (e.g. a bomb) was or was not present in the lower interferometer arm. The novelty of their setup was that the photons propagating through the interferometer sometimes allowed for the answering of the query without interacting with the object in question, i.e. counterfactually.\cite{bPenrose94} Kwiat et al. then showed how the efficiency of this scheme could be taken arbitrarily close to unity by \drma{utilising} a chain of several MZIs.\cite{Kwiat95, Kwiat99}

During the last decade there has been further investigations of counterfactual schemes. Many of these are based on the use of so called ``nested" Mach-Zehnder \drma{i}nterferometers (NMZIs). There have been suggestions that quantum computation \cite{Hosten06}, direct communication \cite{Salih13, Cao17} or transmission of quantum states \cite{Li15} can be conducted without the interrogating particle ever entering the quantum computer in the former case or interacting with the information transmitter in the two latter. However, these schemes have been under intense debate.\cite{Vaidman13, Li13, Vaidman13-2, Salih13, Vaidman14, Salih14, Li15, Vaidman16, Li16, Griffiths16} The criticism resulted in the development of another counterfactual communication (CFC) scheme.\cite{ArvShukur16} The definition of counterfactuality in the former schemes does not allow any particles to travel between receiver and transmitter. Alternatively, the definition in the latter allows particles to travel from receiver to transmitter but not vice versa (i.e. particles are only allowed to propagate in the opposite direction to the message). The essence of the criticism of the former schemes is based on an investigation of the \textit{weak trace} \cite{Vaidman13} that the interrogating particle leaves within the inner part of a NMZI. Essentially, the inner part of the NMZIs have to be inaccessible to the interrogating particle. However, if a weak interaction is present in those parts, the effect of that is of the same order or stronger than if the particle had freely propagated through the interaction.\cite{Vaidman13-2, Vaidman14} The common rebuttal is that this criticism is invalid as a weak interaction alters the perfect interference of the interferometer used in the suggested schemes. 

Whilst the weak trace is an interesting concept, we feel that an argument based on information theoretical principles is desirable to bring clarity to the subtleties of the counterfactual protocols. Furthermore, owing to the intense discussion regarding NMZIs, we also see the need of a thorough investigation of parameter estimation in these structures.

In this paper we adapt the information measures of phase estimation such that they can be used for parameter estimations in Mach-Zehnder interferometer structures. We firstly give an outline of the theoretical framework of the paper. Secondly, we provide a detailed analysis of the wavefunction evolution through the NMZI devices. We observe how the classical Fisher information and  Shannon mutual information changes between experiments with NMZIs depending on where in the interferometer an interacting medium is placed. Furthermore, we evaluate the two different definitions of CFC, with suitable \drma{operational (and interpretation independent)} measures of their respective violations.  The underlying argument of our work, is that absolutely lossless and pure spatial transmission of quantum particles is not attainable. Thus, a theory that relies on such perfect quantum channels is as valid (read: invalid) as a thermodynamic proof only valid at $0$ K. We provide a model of realistic devices that contain a weak un-controlled polarization rotation. This serves the purpose of mimicking real quantum evolutions. Based on this, we can evaluate the counterfactuality of the communication schemes according to their respective definitions. We see that some ``counterfactual" protocols violate their definition of counterfactuality more than a free-space propagation from a transmitter to a receiver. We can thus rule out the counterfactuality of these schemes.

\section{Information Measures}
\label{Sec:InfMeas}
The basis of this work is the knowledge of how the single particle wavefunction (and thus the probability density distribution) evolves through the devices we wish to investigate. The wavefunction evolution is provided by the calculation of unitary operations on some initial quantum state: $\ket{\psi_{in}} \rightarrow \hat{U} \ket{\psi_{in}}$. Throughout the evolution, information will be encoded in the probability density distribution via the interactions that act on the wavefunction. Whilst the extension to multi-photon states is straightforward, we wish to keep this study in line with suggested counterfactual schemes \cite{Salih13, Hosten06, Li15, ArvShukur16}, and restrict our work to single-photon inputs.

It is often nonsensical to ask where a quantum particle has been present between two observations. However, along the evolution of the quantum state, from input to output, one can introduce an interaction that results in parts of the wavefunction occupying a quantum state that only is made available via this interaction. The wavefunction will carry some \textit{information} about the nature of that interaction and it is possible to interpret parts of the probability density---that occupy states only made available via this interaction---as having had a \textit{presence} at the area where the interaction was located. The probability outputs, at the end of the quantum evolution, allow for the estimation of the interaction parameters. The effectiveness of such an estimation, for a given quantum device, can be evaluated with the two information measures outlined in the following subsection.

\subsection{Shannon Mutual Information \& Fisher Information} \label{ssec:ShanFish}

Consider an experiment, given an input state, $\psi_{in}$, and a parameter, $\theta$, that sets some interaction. We can calculate the Shannon mutual information, $H(\theta : M)$, between the parameter $\theta$ and the measurement outcomes $M = \{ M_i \}$, where $M_i$ represents an event that occurred in the $i^{\rm{th}}$ detector location of the total spatial Hilbert space, $\mathcal{H}$:
\begin{align}
H(\theta : M) = \sum_{i \in \mathcal{H}} \int_{\theta_{min}}^{\theta_{max}}  \,d\theta P(M_i | \theta, \psi_{in}) p(\theta) \nonumber \\ \times
\log_{2} \Bigg[ \frac{P(M_i | \theta, \psi_{in})}{\overline{P}(M_i |  \psi_{in})} \Bigg] ,
\label{Eq:Shannon}
\end{align}
where $\overline{P}(M_i |  \psi_{in}) = \int_{\theta_{min}}^{\theta_{max}}  \,d\theta' P(M_i | \theta', \psi_{in}) p(\theta')$, $P(M_i | \theta, \psi_{in})$ is the probability of $M_i$ for some specific $\theta$ and $\psi_{in}$, and $p(\theta)$ is the \textit{a priori} probability distribution of the parameter $\theta$ \cite{bCover06, Bahder06}. The Shannon mutual information provides a measure of how much information about $\theta$ that can be obtained through knowledge of the measurement outcomes of a specific experiment. A large value of $H(\theta : M)$ indicates a good device for parameter estimations of an unknown parameter $\theta$. 

The Shannon mutual information takes into account a prior distribution of $\theta$: $p(\theta)$. However, if the value of $\theta$ is fixed but unknown, one might ask oneself how much information, on average, a single use of a specific interferometer yields about $\theta$. This quantity of information is given by the classical Fisher information \cite{bCover06, Bahder11}:
\begin{equation}
F (\theta)=\sum_{i \in \mathcal{H}} \frac{1}{P(M_i | \theta, \psi_{in}) } \Bigg[ \frac{\partial}{\partial \theta} P(M_i | \theta, \psi_{in}) \bigg]^2 .
\label{Eq:Fisher}
\end{equation}
In the Cramér-Rao inequality $F (\theta)$ sets a lower bound on the variance of the estimator of $\theta$, $\theta_e$, obtained in a specific experiment:
\begin{equation}
\rm{Var}(\theta_e) \geq \frac{1}{F(\theta)} .
\end{equation}

\subsection{Fisher Information as a Measure of Presence in Optical Circuits}

\begin{figure}
\centering
\includegraphics[scale=0.24]{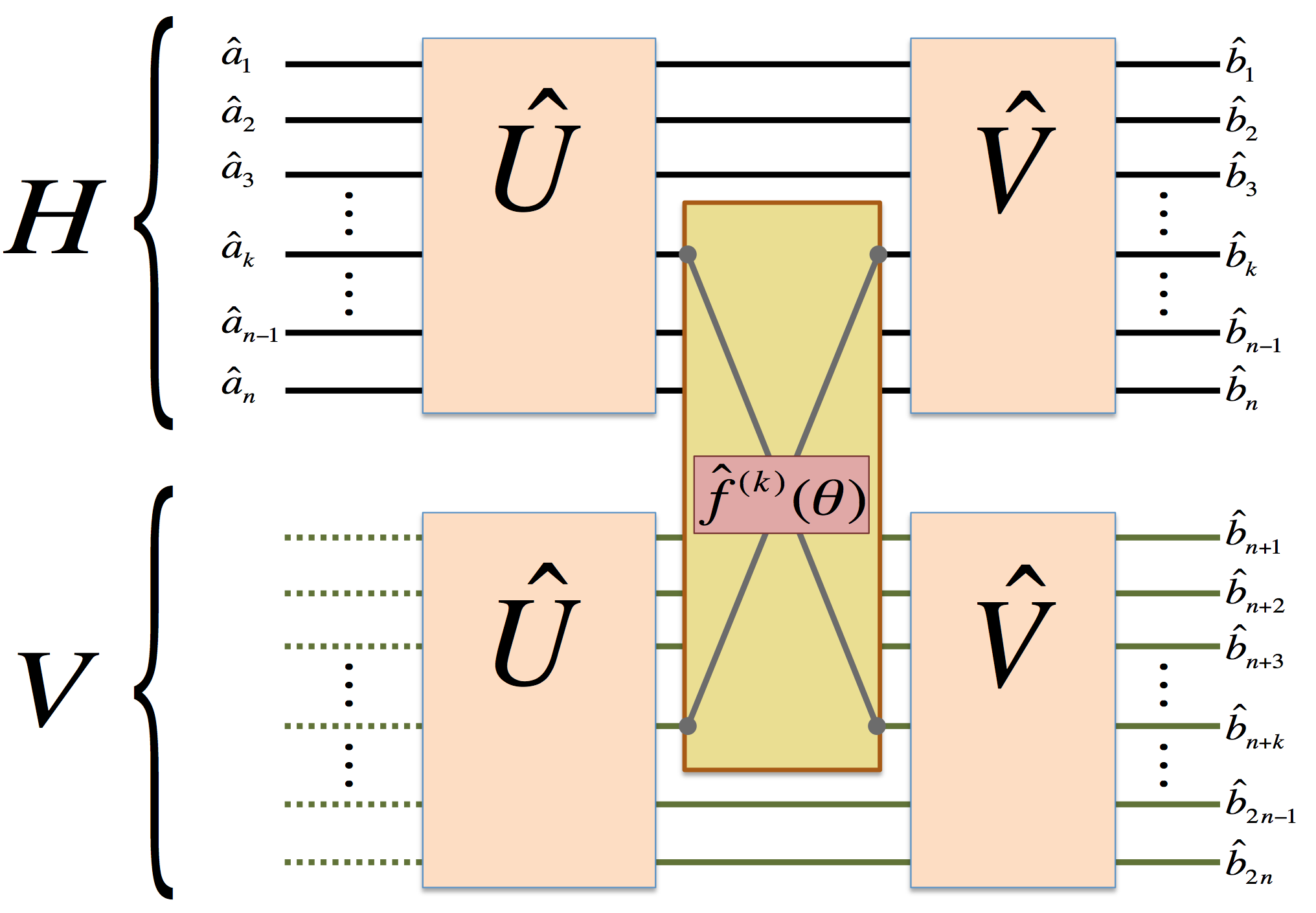}
\caption{(color online) Sketch of an optical circuit of the form of Eq. \ref{Eq:Scatter}, which is described in the text. }
\label{fig:OptFish}
\end{figure}

Many optical quantum interferometers do not involve polarization operations. However, any ``real" experiment with single photons will naturally include some polarization operations---owing to, for example, material impurities and systematic errors in the experimental setup. We mimic the inevitable imperfect nature of \textit{real} quantum channels by introducing single polarization interactions somewhere in the optical circuits.\footnote{Weak unwanted polarization interactions can occur in all parts of a realistic device. However, our setup can be considered as one where the experimentalist, Alice, is allowed to correct for unwanted polarization rotations from all parts of the device, except for those that occur in a location controlled by Bob (central rectangle in Fig. \ref{fig:OptFish}). Hence, we consider one unwanted polarization rotation, due to Bob's laboratory.} \drma{This represents an interaction with Bob's laboratory rather than a generic noise model.} We call this interaction the ``tagging" of the wavefunction. By introducing the polarization degree of freedom, we can use the Fisher information to estimate the parameters associated with the interaction. As has been described above, a tagged part of the wavefunction can be considered to have \drma{previously existed} at the location of the polarization interaction. In this subsection we show that, given access to all the outcome possibilities, the classical Fisher information, in the interferometers studied in this work, is always proportional to the integrated probability density distribution that has evolved through the interaction in the Schr\"odinger picture. 

Firstly, we define an optical input vector, $\bm{a}$, of length $2n$, which evolves into an output vector, $\bm{b}$. The $2n$ levels correspond to the $n$ different optical paths of the device, each of which can exist in one of the two polarization states. We choose the order of the vector elements so that the first $n$ entries have the polarization of the initial input state, and so that the following $n$ entries have orthogonal polarization.

We can describe the evolution of the input state, $\bm{a}$, through the interferometer by a scattering matrix, $\mathcal{S}$, in terms of three operations:

\begin{equation}
\label{Eq:Scatter}
\mathcal{S} \bm{a} \equiv \big(\hat{V} \cdot \hat{f}^{(k)}(\theta) \cdot \hat{U} \big) \bm{a} \equiv \bm{b}^{(k)} ,
\end{equation}
where $\hat{U}$ and $\hat{V}$ are the unitary operators of the evolution up to and after the tagging polarization rotation respectively, and $\hat{f}^{(k)}(\theta)$ is the unitary operator that describes the \textit{single} polarization rotation at the specific spatial path of $k$ (where $1 \leq k \leq n$) by an angle $\theta$. A sketch of the optical circuit of $\mathcal{S}$ is given in Fig. \ref{fig:OptFish}. Note that  $\hat{U}$ and $\hat{V}$ acts solely on the spatial degree of freedom and do not manipulate the polarization of the wavefunction.

We can define the wavefunction after $\hat{U}$ has been applied as $\bm{c} \equiv \hat{U} \bm{a}$, with:
\begin{align}
\label{Eq:c}
c_i = \sum_{j=1}^{2n} U_{i,j} a_j .
\end{align}

The  rotation matrix is then applied to this state. 
It rotates the quantum state between two of the vector levels, $k$ and $k^{\prime}$. It can be represented by the following real matrix:
\begin{align}
\label{Eq:RotMatrx}
\hat{f}^{(k)}(\theta)  = &
  \begin{pmatrix}
    1 & \cdots & 0 & \cdots & 0 & \cdots & 0\\
    \vdots & \ddots & \vdots & \ddots & \vdots & \ddots & \vdots  \\
    0 & \cdots & f_{k,k}(\theta) & \cdots & f_{k,k^{\prime}}(\theta)& \cdots & 0 \\
   \vdots & \ddots & \vdots & \ddots & \vdots & \ddots & \vdots  \\
    0 & \cdots & f_{k^{\prime}, k}(\theta) & \cdots & f_{k^{\prime}, k^{\prime}}(\theta)& \cdots & 0 \\
    \vdots & \ddots & \vdots & \ddots & \vdots & \ddots & \vdots  \\
     0 & \cdots & 0 & \cdots & 0 & \cdots & 1
  \end{pmatrix} ,
\end{align}
where $f_{k, k}(\theta) = \sqrt{1- \big(f_{k,k^{\prime}}(\theta)\big)^2}$, $f_{k^{\prime}, k}(\theta) = - f_{k, k^{\prime}}(\theta)$ and $f_{k^{\prime}, k^{\prime}}(\theta) = f_{k,k}(\theta)$.

We further arrange the vector entries such that $l$ and $l+n$ correspond to the same spatial location for $1 \leq l \leq n$. This means that $k^{\prime} = k + n$ in Eq. \ref{Eq:RotMatrx} and $c_{i} = 0$ for $i>n$ in Eq. \ref{Eq:c}. We express the quantum state after the polarization interaction as $\bm{d}^{(k)}(\theta) \equiv \hat{f}^{(k)}(\theta) \bm{c}$, with:
\begin{align}
\label{Eq:d}
d_i^{(k)}(\theta) = \sum_{j=1}^{2n} f_{i,j}^{(k)}(\theta) c_j .
\end{align}
We note that the only components of $\bm{d}^{(k)}$ that depend on $\theta$ are $d_{i=k}^{(k)} = f_{k,k}(\theta) c_k$ and $d_{i=k+n}^{(k)} = f_{k+n,k}(\theta) c_k$. 

Finally, we can apply the last unitary evolution $\hat{V}$. Following the steps above, we express the output vector as $\bm{b}^{(k)}(\theta) \equiv \hat{V}\bm{d}^{(k)}(\theta)$, with:
\begin{align}
b_i^{(k)}(\theta) \equiv \beta^{(k)}_i + b^{(k)}_{i,\theta}(\theta) = \sum_{j=1}^{2n} V_{i,j} d_j^{(k)}(\theta) ,
\end{align}
where the $\theta$-dependence of $b_i^{(k)}(\theta)$ is encapsulated in $b_{i,\theta}^{(k)}(\theta)$ and a corresponding term, independent of $\theta$, is defined as $\beta^{(k)}_i$.

The probability of measuring the single-photon in the $i^{\rm{th}}$ output port, is then given by:
\begin{equation}
P_i^{(k)} (\theta) = |\beta_i^{(k)} + b_{i, \theta}^{(k)}(\theta)|^2 .
\end{equation}
This can be re-expressed as 
\begin{equation}
P_i^{(k)} (\theta) \equiv |\beta_i^{(k)} + b_{i}^{\prime(k)} f_{j,k}^{(k)}(\theta)|^2 ,
\end{equation}
where $j=k$ if $i\leq n$ and $j=k+n$ if $i>n$. For convenience we now drop the $(k)$ superscript.

Using Eq. \ref{Eq:Fisher}, the individual Fisher information components can be expressed as:
\begin{align}
F_i  = & \frac{1}{\big| \beta_i + b_{i}^{\prime}f_{j,k}(\theta) \big|^2}  \Big[\frac{\partial}{\partial \theta} \big|\beta_i + b_{i}^{\prime}f_{j,k}(\theta)\big|^2\Big]^2 \nonumber \\
     = & \frac{1}{\big( \beta_i + b_{i}^{\prime}f_{j,k}(\theta) \big)\big( \beta_i^{ \ast} + b_{i}^{\prime\ast}f_{j,k}(\theta) \big)} \nonumber \\ 
     \times & \Big[\frac{\partial}{\partial \theta} \big(\beta_i + b_{i}^{\prime}f_{j,k}(\theta)\big)\big(\beta_i^{ \ast} + b_{i}^{\prime \ast}f_{j,k}(\theta)\big)\Big]^2 \nonumber \\
\end{align}
This expression can be simplified by expressing the coefficients as $\beta_i \equiv | \beta_i | e^{i\phi_{i}}$ and $b_{i}^{\prime} \equiv | b_{i}^{\prime} | e^{i\phi_{i,\theta}}$ and defining $\Phi_i \equiv \phi_{i}-\phi_{i,\theta}$:
\begin{align}
F_i    = & 
     \frac{ \big(   \cos{(\Phi_i)} |\beta_i|  +   |b_{i}^{\prime}| f_{j,k}(\theta)   \big)^2}
     {\big( |\beta_i|^2 + |b_{i}^{\prime}|^2 f_{j,k}^2(\theta) + 2\cos{(\Phi_i)|\beta_i||b_{i}^{\prime}|f_{j,k}(\theta)} \big)} \nonumber \\
     \times &
     4 |b_{i}^{\prime}|^2 \Big( \frac{\partial}{\partial \theta} f_{j,k}(\theta)   \Big)^2 . \nonumber \\
\end{align}

We notice that if the phase difference $\Phi_i$ is a multiple of $\pi$, the expression simplifies to:
\begin{align}
F_i   = & 4 |b_{i}^{\prime}|^2 \Big( \frac{\partial}{\partial \theta} f_{j,k}(\theta)   \Big)^2 . \nonumber \\
\end{align}
This phase criterion is satisfied for all $i$ if the phases of all the spatial comontents, $i \leq n$, of the input state are the same and $\mathcal{S}$ is real (e.g. the optical setup only contains beam-splitters that can be represented by real operators). It is also satisfied by all the optical setups considered in Refs. \cite{Elitzur93, Kwiat95, Kwiat99, Vaidman13, Salih13, ArvShukur16}. 

For the quantum optical setups of interest in this paper, we can thus express the classical Fisher information (Eq. \ref{Eq:Fisher}) as:
\begin{equation}
F(\theta)  = \sum_{i=1}^{2n} 4   \Big[ \frac{\partial}{\partial \theta} \big| b_{i, \theta}(\theta) \big| \Big]^2 .
\end{equation}

This can be re-expressed as:
\begin{align}
F(\theta)  = &  \sum_{i=1}^{n} 4   \Big[ \frac{\partial}{\partial \theta}\big|V_{i, k}d_k(\theta) \big|\Big]^2  \nonumber \\ 
+ & \sum_{i=n+1}^{2n} 4   \Big[ \frac{\partial}{\partial \theta}\big|V_{i, k+n}d_{k+n}(\theta) \big|\Big]^2  \nonumber \\ 
= &  \sum_{i=1}^{n} 4 \big|V_{i, k}\big|^2  \Big[ \frac{\partial}{\partial \theta}\big|d_k(\theta) \big|\Big]^2  \nonumber \\ 
+ & \sum_{i=n+1}^{2n} 4  \big|V_{i, k+n}\big|^2 \Big[ \frac{\partial}{\partial \theta}\big|d_{k+n}(\theta) \big|\Big]^2
\end{align}
As $\hat{U}$ and $\hat{V}$ do not manipulate the polarization of the wavefunction, the symmetry of the matrix $\hat{V}$ is such that $V_{i, k}=V_{i+n, k+n}$. It also implies that $V_{i, k}=0$ for $i>n$ and $V_{i, k+n}=0$ for $i \leq n$ . By assuming a real $\mathcal{S}$ and defining a suitable reference-point for the global input phase we simplify our expression further:
\begin{align}
F(\theta)  = &  \sum_{i=1}^{2n} 4   |V_{i, k}|^2   \Big[ \frac{\partial}{\partial \theta} |d_k(\theta)|  \Big]^2    \nonumber \\ 
+ &
 \sum_{i =1}^{2n} 4 |V_{i, k}|^2   \Big[ \frac{\partial}{\partial \theta} |d_{k+n}(\theta)| \Big]^2 .
\end{align}
We sum the squared entries in the column of our unitary matrix to unity, and the expression simplifies to:
\begin{align}
F(\theta)  = & 4  \Big[  \frac{\partial}{\partial \theta}  |d_k(\theta)  |\Big]^2 
+ 
 4   \Big[ \frac{\partial}{\partial \theta} |d_{k+n}(\theta) |\Big]^2 .
\end{align}

At this stage we note that the total Fisher information of the device does not contain any dependence on the unitary operation $\hat{V}$. 

We continue by substituting the expressions of $d_{k}$ and $d_{k+n}$ from above (Eq. \ref{Eq:d}) to obtain a final expression of the Fisher information:
\begin{align}
\label{Eq:OptFish}
F^{(k)}(\theta)  = & 4  |c_k|^2 \Big( \Big[  \frac{\partial}{\partial \theta}  \big(f_{k,k}(\theta)  \big)\Big]^2 
+ 
    \Big[ \frac{\partial}{\partial \theta} \big(f_{k+n,k}(\theta) \big)\Big]^2\Big) \nonumber \\
    = & 4  |c_k|^2 \frac{\Big( \frac{\partial}{\partial \theta} f_{k+n,k}(\theta)    \Big)^2}{\big(f_{k,k}(\theta)\big)^2},
\end{align}
where we briefly re-introduce the $(k)$ superscript. 

To conclude this section, we make the observation that the Fisher information, Eq. \ref{Eq:OptFish}, is proportional to $|c_k|^2$. $|c_k|^2$ is the probability of observing the photon in the $k^{\text{th}}$ spatial state if a detector had been placed at the location of the tagging interaction. Na\"ively the extent to which the wavefunction spreads into a spatial location, according to the time-dependent Schr\"odinger equation, could be interpreted as a measure of how much the particle has been \textit{present} there. However, owing to the nature of quantum mechanics, it is philosophically problematic to specify what the physical meaning of the wavefunction between measurements is.\footnote{This is discussed at length in Wheeler's ``The `Past' and the `Delayed-Choice' Double-Slit Experiment", which has been reproduced in ref. \cite{bMarlow78}} Nevertheless, in the circuits studied above, the tagging mechanism, which rotates the initial polarization of the photon into a superposition state, is the only polarization component of the interferometer. Hence, it is arguably less na\"ive to consider an output photon in an altered polarization state to have had a past that has included the passage through the tagging part of the interferometer. Even if the introduced tagging polarization rotations are \last{vanishingly} small and do not affect the specific interferometer significantly, Eq. \ref{Eq:OptFish} shows that the information content, which travels from the polarization rotator to the output ports, is weighted by the square of the integrated wavefunction at the location of the interaction. Hence, the Fisher information is arguably a good measure of the \textit{extent} to which a particle can be considered to previously have had a presence at the tagging location.

\color{black}
We finish this section with a note regarding an extension of the above theory, to include other degrees of freedom. If an interferometer contains polarization rotations (such as the Michelson-based device in Ref. \cite{Salih13}), these rotations can simply be included in $\hat{U}$ and $\hat{V}$. The single rotation, that we calculate the Fisher information with respect to, should then be changed to an alternative degree of freedom. With the corresponding alterations of the measure, $\hat{f}^{(k)}(\theta)$ can, for example, be taken to be a weak rotation of the photon's internal orbital angular momentum. The beauty of this analysis is that Eq. \ref{Eq:OptFish} will still be valid. However, as the interferometers considered
in this paper are all ideally non-polarizing, we conduct our analysis using a weak disturbances on the polarization.
\color{black}

\section{Measures of Counterfactual Violation}

In this section we develop measures for the extent a process violates counterfactuality. \drma{These measures will be used in Section \ref{Sec:Type1Vio} and \ref{Sec:Type2Vio} to investigate counterfactual violations of the CFC protocols proposed by Salih et al. \cite{Salih13} and Arvidsson-Shukur and Barnes \cite{ArvShukur16}.}

\subsection{Two CFC Definitions}

\begin{figure}
\centering
\includegraphics[scale=0.28]{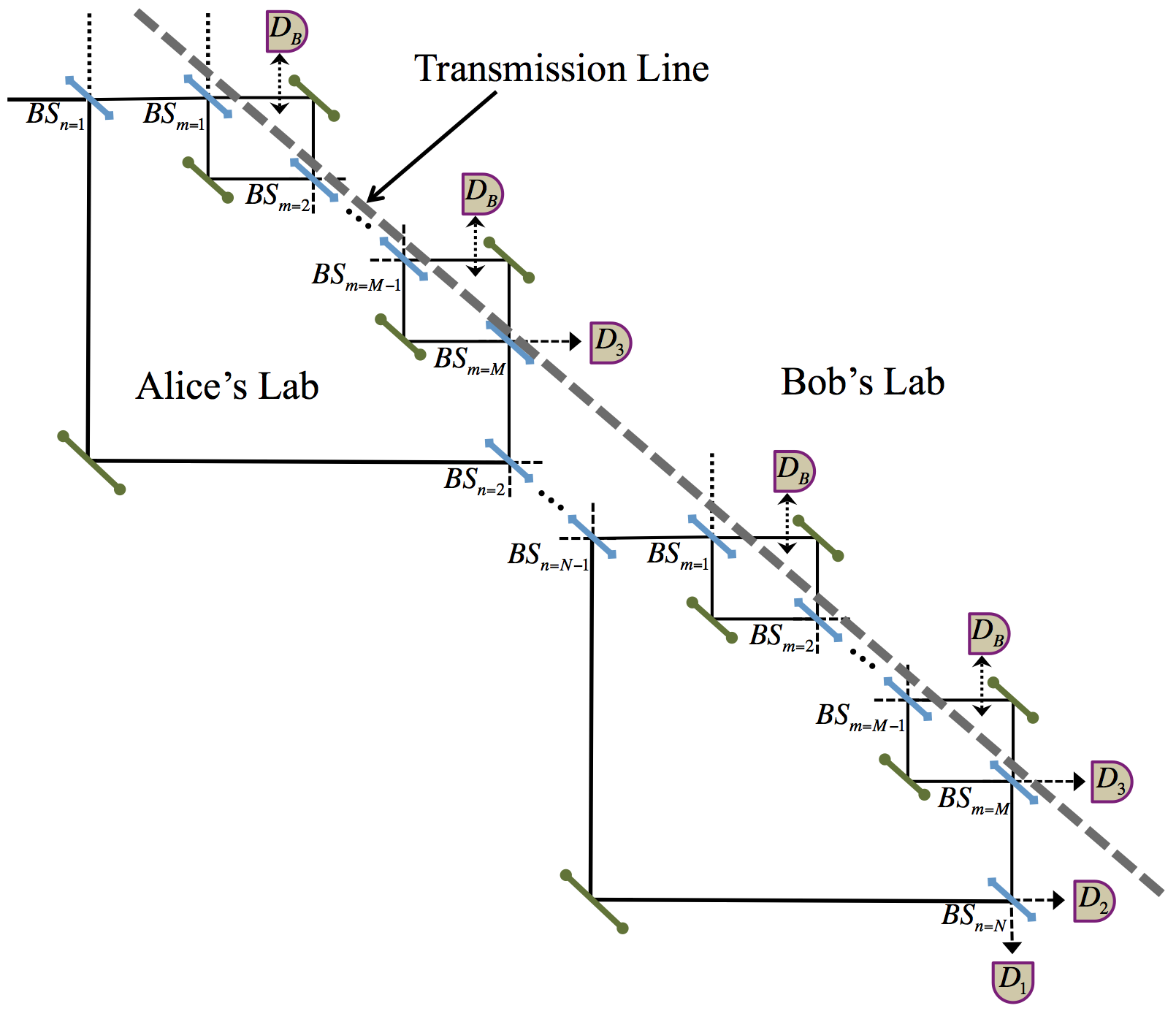}
\caption{(color online) The chained nested Mach-Zehnder interferometer suggested for CFC in Ref. \cite{Salih13}. Alice inputs a photon in the upper left path and Bob has the choice of introducing detectors in his part of the device. His choice governs the statistics of the final output detections at $D_1$ and $D_2$.}
\label{fig:ChainedNMZI}
\end{figure}

\begin{figure}
\centering
\includegraphics[scale=0.32]{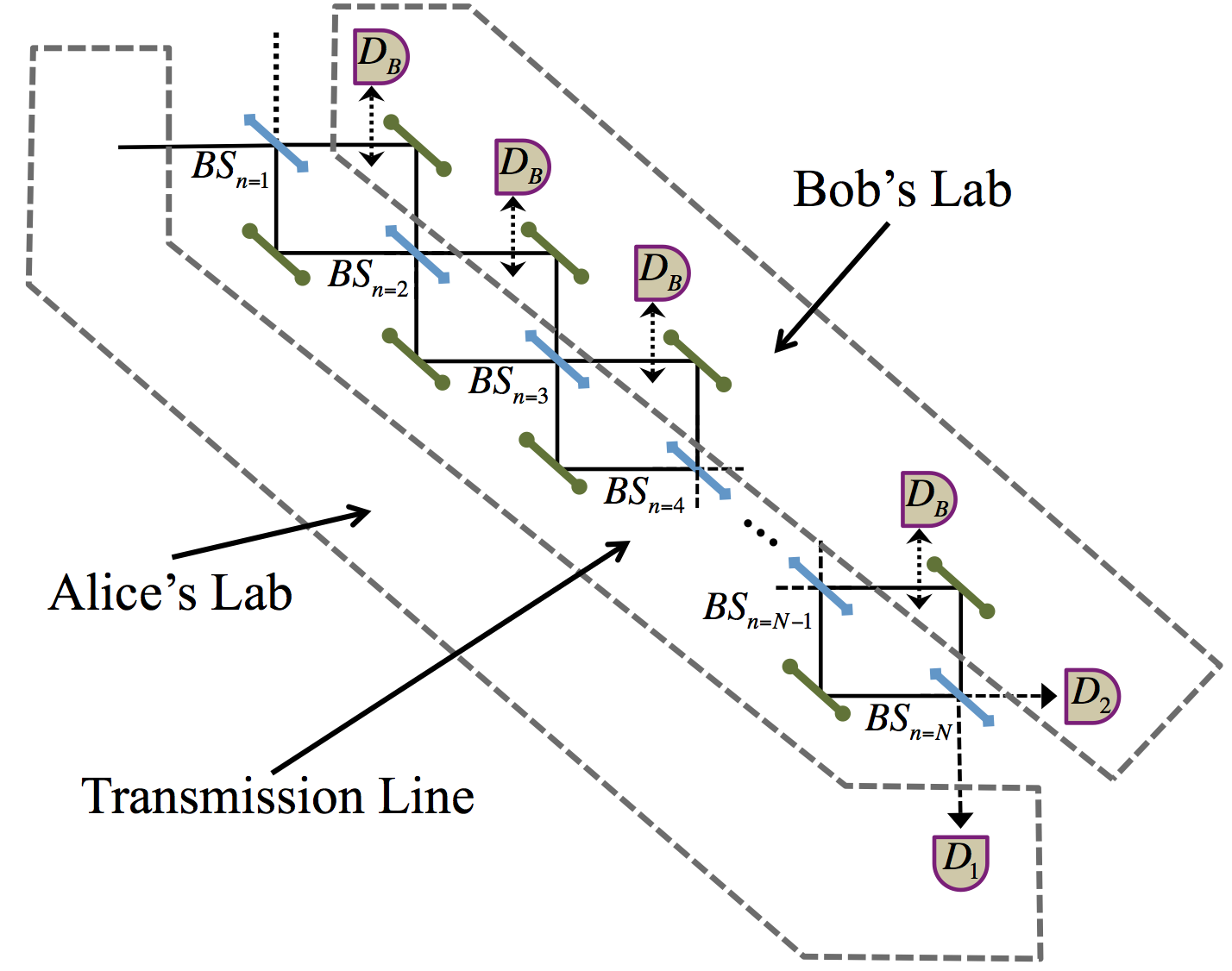}
\caption{(color online) The chained Mach-Zehnder interferometer as used in the CFC scheme of Ref. \cite{ArvShukur16}. \drma{ As in Fig. \ref{fig:ChainedNMZI}, Alice inputs a photon in the upper left path and Bob governs the statistics of the final output detections at $D_1$ and $D_2$ by inserting or not inserting detectors respectively.}}
\label{fig:CMZI}
\end{figure}

There are two main schemes for CFC, based on different definitions of the concept. One was developed in 2013 by Salih et al.\cite{Salih13} (See Fig. \ref{fig:ChainedNMZI}). We refer to the definition of CFC in that protocol as the Type \rom{1} definition. Another scheme was developed by Arvidsson-Shukur and Barnes in 2016 \cite{ArvShukur16} (see Fig. \ref{fig:CMZI}) and we refer to its definitions as the Type \rom{2} definition. 

\last{In both schemes Bob transmits a message to Alice. The schemes respective bit-transmissions are initiated by Alice sending a single particle into the upper left input path of the devices. Bob can then choose to transmit a $0$ bit or a $1$ bit by keeping his laboratory free or inserting absorbing detectors respectively. In the Type \rom{1} protocol (Fig. \ref{fig:ChainedNMZI}), the quantum Zeno effect \cite{Degasperis74, Misra77} is used for both bit-processes, such that the particle ends up at Alice's detectors $D_1$ ($0$ bit) or $D_2$ ($1$ bit) without ever having crossed the transmission line between Alice and Bob. The Type \rom{2} protocol (Fig. \ref{fig:CMZI}), on the other hand, only utilise the quantum Zeno effect for the $1$ bit process. In this process the particle enters the transmission line and returns to Alice, but it never enters Bob's laboratory. In the $0$ bit process the particle travels from Alice's laboratory into the transmission line. It then evolves into Bob's laboratory. The protocols are described in further detail in Section \ref{Sec:Type1Vio} and \ref{Sec:Type2Vio} of this paper. The differences in the counterfactuality definitions are summarised in Table \ref{Table:CountFac}.}

\begin{table}
\caption{\label{Table:CountFac} The counterfactual nature of the Type \rom{1} \cite{Salih13} and the Type \rom{2} \cite{ArvShukur16} communication schemes. In both schemes a message is transmitted from a transmitter, Bob, to a receiver, Alice.}
\begin{tabular}{ |p{2cm}||p{2.8cm}|p{2.8cm}|  }
 \hline
    Scheme $\setminus$ Bit & Logical 0: & Logical 1:\\
 \hline  \hline
  Type \rom{1}: & \multicolumn{2}{l|}{No particles cross the transmission line} \\
     & \multicolumn{2}{l|}{between Alice and Bob.} \\
  \hline
  Type \rom{2}:   &  Particles propagate from Alice to Bob via the transmission line. & Particles propagate from Alice to the transmission line and back again.\\

 \hline
\end{tabular}
\end{table}

The counterfactuality of an interaction-free process, naturally depends on the boundaries of the spatial extent of the respective ``laboratories" of the participants in the protocol. Both the Type \rom{1} and \rom{2} definitions state that for a process in Bob's laboratory to be counterfactual with respect to Alice, it is essential that particles should never propagate from Bob to Alice (such that parts of a wavefunction that interacts with Bob's laboratory will have a vanishing probability to end up in Alice's). 

\last{As mentioned above,} the Type \rom{1} and \rom{2} schemes utilise the quantum Zeno effect, \drma{triggered by absorbing detectors in Bob's laboratory}, to produce their respective logical $1$ bit-values in a counterfactual manner. We are not aware of any works disputing the counterfactual nature of these processes.  That leaves us with the task of evaluating the Type \rom{1} and \rom{2} logical $0$ processes. 

\subsection{The Measures}
The Type \rom{1} definition should forbid particles to propagate from Alice to Bob and vice versa. Following the discussion in the section above, a good measure of the violation of such a process could be based on the Fisher information encoded at Bob's laboratory, in a particle originating from Alice.

In order to evaluate the ``strength" of a violation of a Type \rom{1} logical $0$ process (caused by a tagging interaction as discussed above) we need a Fisher information benchmark. In this work we benchmark with respect to the Fisher information of a free-space evolution, $F_{free}$, subject to the same tagging interaction as the circuits of interest. This can, for example, be the scenario of Bob directly sending a photon to Alice (see Fig. \ref{fig:SingPhot})---a clearly non-counterfactual scenario. We can now define our new measure for the violation of Type \rom{1} counterfactuality. We call the measure the Type \rom{1} counterfactual violation strength:
\begin{equation}
\label{Eq:CountVio}
\mathcal{D} \coloneqq  \frac{F}{F_{free}} .
\end{equation}
This quantity can effectively be thought of as the Fisher information encoded in a particle originating from Alice, owing to an interaction at Bob's laboratory, as a fraction of the Fisher information of a free-space interaction. A value of $\mathcal{D}=0$ corresponds to no wavefunction interacting with Bob's laboratory; and a value of $\mathcal{D} \geq 1$ corresponds to an interaction stronger than or equal to that of a free-space interaction. Values of the order of unity or bigger are convicted of \drma{fully} violating Type \rom{1} counterfactuality.

When it comes to evaluating the logical $0$ in Type \rom{2} protocols, we need a different measure than Eq. \ref{Eq:CountVio}. This is because this scheme allows Alice's particles to be encoded by Bob, as long as they do not return to Alice.\cite{ArvShukur16} The probability of detection in Alice's laboratory in this process is null for perfect quantum channels ($\theta=0$). Thus, a reasonable measure of a counterfactual violation would be the total probability of a particle returning to Alice as a result of a non-collapsing interaction (i.e. $\theta \neq 0$) in Bob's laboratory. Let $M_{j \in \mathcal{H}_{\bm{A}}}$ denote the measurement outcomes triggered in the Hilbert space of states within the spatial extent of $\bm{A}$. Additionally, let $ M_{j^\prime \notin \mathcal{H}_{\bm{A}}}$ denote the negative measurement that indicate all outcome states outside the spatial extent of $\bm{A}$. Our new Type \rom{2} measure can then be expressed as:

\begin{equation}
\label{Eq:CountVioProb}
P_{\bm{A}} \coloneqq \sum_{j \in \mathcal{H}_{\bm{A}}} P(M_j | \theta, \psi_{in})=1-P(M_{j^\prime \notin \mathcal{H}_{\bm{A}}} | \theta, \psi_{in}) .
\end{equation}
This measure can be interpreted in a way such that $P_{\bm{A}}=1$ corresponds to a full counterfactual violation of the Type \rom{2} logical $0$ and $P_{\bm{A}}=0$ corresponds to perfect counterfactuality.

Moreover, even though the probability to trigger a detection in Alice's laboratory can be very small ($P_{\bm{A}} \approx 0$), small probabilities can generate large Fisher informations.  We thus introduce the spatially restricted Fisher information, $F_{\bm{A}}$. $F_{\bm{A}}$ is a measure of the sum of the individual components of the classical Fisher information (Eq. \ref{Eq:Fisher}), as experienced by an observer $\bm{A}$. We define:
\begin{align}
\label{Eq:FishSpacial}
F_{\bm{A}} (\theta) \coloneqq & \sum_{j \in \mathcal{H}_{\bm{A}}} \frac{1}{P(M_j | \theta, \psi_{in}) } \Bigg[ \frac{\partial}{\partial \theta} P(M_j | \theta, \psi_{in}) \bigg]^2 \nonumber \\
& +  \frac{1}{1 - P_{\bm{A}} } \Bigg[ \frac{\partial}{\partial \theta} P_{\bm{A}} \bigg]^2 .
\end{align}
The first term corresponds to the summation of the Fisher information components of particle detections by $\bm{A}$, whilst the second term corresponds to the Fisher information component of negative measurements by $\bm{A}$, i.e. when no particle is detected by $\bm{A}$.
Interestingly, in a Type \rom{2} scheme, Alice can still obtain a large Fisher information of Bob's $\theta$, even if counterfactuality is only violated weakly by $\theta$.

\drma{
We are now in the position to evaluate  Type \rom{1} and Type \rom{2} CFC protocols. However, first we conduct an elaborate study of parameter estimation in NMZIs. This is done in the next section, where we utilise the information measures from Section \rom{2}.\rom{1} and extend the works of Bahder \textit{et al.} \cite{Bahder06, Bahder11} to NMZI structures.  In Section \ref{Sec:Type1Vio} and \ref{Sec:Type2Vio}, we then expand the analysis in order to evaluate the counterfactuality of Type \rom{1} and \rom{2} CFC protocols.
}

\section{Information in Nested MZIs} \label{Sec:InfNMZI}

\subsection{Free Space Interaction}
\begin{figure}
\centering
\includegraphics[scale=0.50]{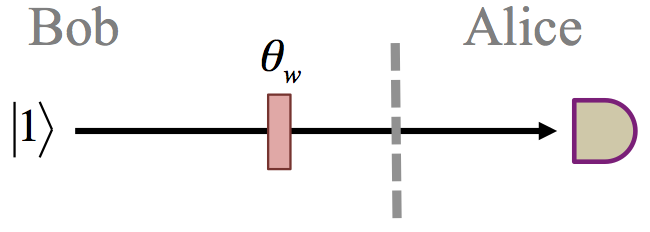}
\caption{(color online) A single-photon state is incident on a weak polarization rotator. The photon is then measured in its number state and polarization state.}
\label{fig:SingPhot}
\end{figure}

As a reference scenario for the optical circuits discussed in following sections of this paper, we provide the simplest of examples of perturbations caused by a polarization rotator. We consider a single-photon state. It has a polarization degree of freedom, and propagates in a straight line. It interacts with a polarizing medium---the rotator---shortly after which it is measured. (See Fig. \ref{fig:SingPhot}).

The quantum interaction of the wavepacket with the rotator results in a rotation of the polarization set by the parameter $\theta = \theta_w$. We can calculate the \drma{Fisher information, $F(\theta_w)$, and the} Shannon mutual information, $H(\theta_w : M)$, of the parameter $\theta_w$ and the measurement outcomes.

 We introduce the creation operators $\hat{a}_H^{\dagger}$ and $\hat{a}_V^{\dagger}$, which create a single photon in a horizontal and vertical state respectively. Our polarization axes are defined such that the input state can be written as:
\begin{equation}
\hat{a}_H^{\dagger} \ket{0} \equiv 
\begin{pmatrix}
   1  \\
  0 
 \end{pmatrix} .
\end{equation}
For the case of a single polarizing rotator, the scattering matrix (Eq. \ref{Eq:Scatter}) takes the form:

\begin{equation}
\label{Eq:Rot}
\mathcal{S} = \hat{f}^{(1)}(\theta_w)=
\begin{pmatrix}
   \sqrt{1-\theta_w^2} & \theta_w \\
  -\theta_w  & \sqrt{1-\theta_w^2}
 \end{pmatrix} .
\end{equation}
The polarization rotations of the different optical circuits studied in this paper will all be in the form of Eq. \ref{Eq:Rot}. 

The detector in Fig. \ref{fig:SingPhot} measures in the basis: $\ket{n_H, n_V}$, where $n_H$ and $n_V$ are the respective photon numbers of horizontal and vertical polarization at the output. The output probabilities are  given by:
\begin{align}
& P(n_H=1| \theta_w) = 1-\theta_w^2 , \\
& P( n_V=1 | \theta_w) = \theta_w^2 ,
\end{align}
where we have adopted a notation such that the statement $n_s = 1$ implicitly assumes that all other possible measurement outcomes, $t \neq s$, satisfy $\sum_{t \neq s}n_t=0$. As we continue to work with single-photon input states, we keep this notation throughout the paper.

\drma{If $\theta_w$ is fixed,} the classical Fisher information of the free-space rotation is given by Eq. \ref{Eq:Fisher} or Eq. \ref{Eq:OptFish}:
\begin{equation}
\label{Eq:FreeSpace}
F_{free}=\frac{4}{1-\theta_w^2} .
\end{equation}
Eq. \ref{Eq:FreeSpace} will be used as the free-space benchmarking Fisher information in Eq. \ref{Eq:CountVio} when calculating counterfactual violation strengths further on in this paper.

\drma{If we instead} assume no prior knowledge of $\theta_w$ such that $\theta_{min}=-1$, $\theta_{max}=1$ and $p(\theta) = 1 / 2$ in Eq. \ref{Eq:Shannon}, the mutual information is given by:
\begin{equation}
H(\theta_w : M) =  \frac{\ln(108)-4}{3\ln(2)} \approx 0.328 .
\end{equation}

\subsection{Nested MZI Interaction}

\begin{figure}
\centering
\includegraphics[scale=0.4]{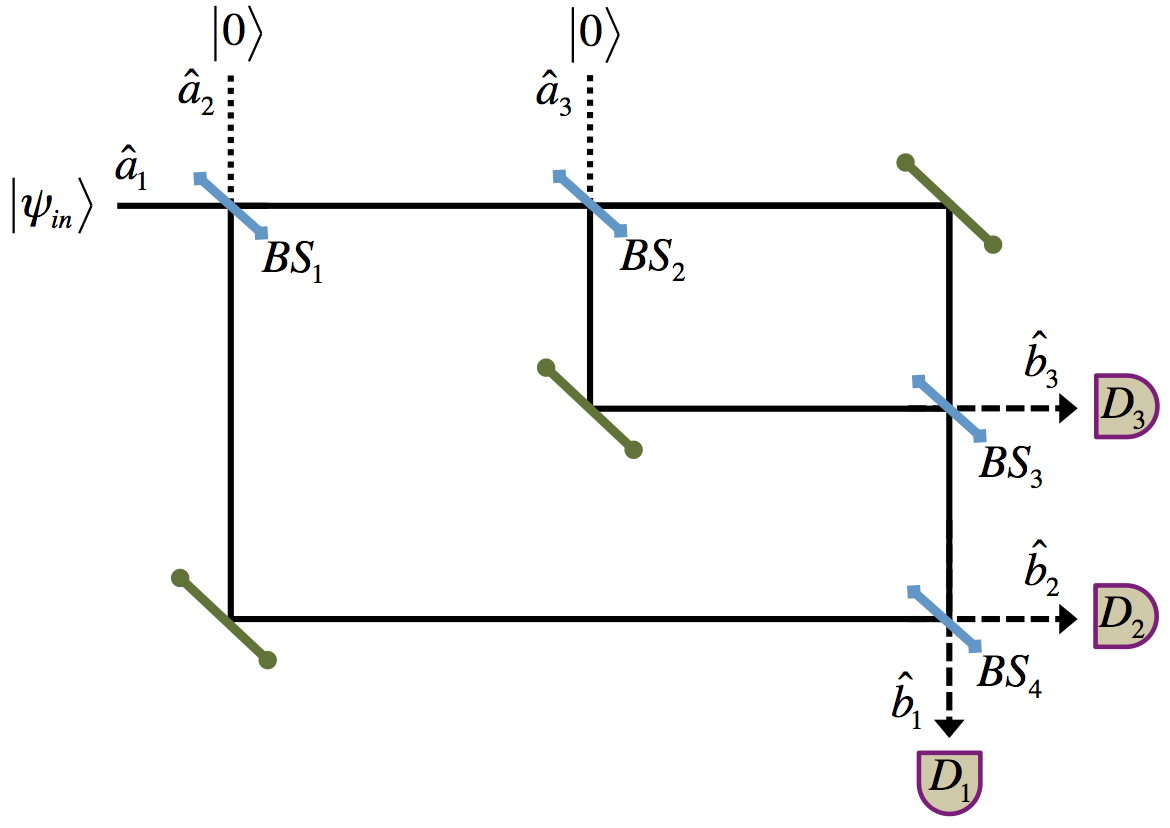}
\caption{(color online) A single-photon state is sent as the input to one of the ports of a NMZI. There are six possible detection outcomes (three spatial outcomes, each having a polarization degree of freedom). The green barred lines and blue arrowed lines represent mirrors and non-polarizing beam-splitters respectively.}
\label{fig:NestMZI}
\end{figure}

We now investigate how the position of a polarization rotator in the nested Mach-Zehnder interferometer (see Fig. \ref{fig:NestMZI}) changes the output probabilities, the \drma{Fisher information and the Shannon mutual information .}

In general one can describe the normalised input and output vectors, $\bm{a}$ and $\bm{b}$, of the NMZIs by:
\begin{align}
\bm{a} = \frac{1}{\sqrt{L_a}}
\begin{pmatrix}
   n_{1,H}^a  \\
   n_{2,H}^a  \\
   n_{3,H}^a  \\
   n_{1,V}^a  \\
   n_{2,V}^a  \\
   n_{3,V}^a 
 \end{pmatrix} , 
 \bm{b}= \frac{1}{\sqrt{L_b}}
\begin{pmatrix}
   n_{1,H}^b  \\
   n_{2,H}^b  \\
   n_{3,H}^b  \\
   n_{1,V}^b  \\
   n_{2,V}^b  \\
   n_{3,V}^b 
 \end{pmatrix} ,
\end{align}
where $L_a$ and $L_b$ are some normalisation constants. The corresponding input creation operators are given by $\hat{a}_i^\dagger$. These are transformed into the output operators via the scattering matrix (see Eq. \ref{Eq:Scatter}) $\hat{b}_j^\dagger=\mathcal{S}_{i,j}\hat{a}_i^\dagger$. In the following sections we restrict our input states to horizontal single-photon states, initially occupying the first spatial input port, such that $\ket{\psi_{in}}=\hat{a}_{1,H}^\dagger\ket{0}$. We drop the superscripts of the vector elements.

The beam-splitters, $BS_i$, have reflection and transmission coefficients $r_i$ and $t_i$ respectively. In the NMZI device $r_2=t_2=r_3=t_3=\frac{1}{\sqrt{2}}$, such that the scattering matrix of the NMZI device, without polarization rotators (i.e. $\hat{f}^{(k)}(\theta)=\hat{1}$ in Eq. \ref{Eq:Scatter}), is given by:

\begin{align}
\mathcal{S} =
\begin{pmatrix}
   r_1 r_4 & t_1 r_4 & t_4 & 0 & 0 & 0 \\
   -r_1 t_4 & -t_1 t_4 & r_4 & 0 & 0 & 0 \\
   t_1 & -r_1 & 0 & 0 & 0 & 0 \\
   0 & 0 & 0 & r_1 r_4 & t_1 r_4 & t_4 \\
   0 & 0 & 0 & -r_1 t_4 & -t_1 t_4 & r_4 \\
   0 & 0 & 0 & t_1 & -r_1 & 0 
 \end{pmatrix} .
\end{align}

We proceed by evaluating the NMZI device for parameter estimation by considering five different scenarios with a polarization rotator placed in one out of five locations in the NMZI device. (See Fig. \ref{fig:NestMZIOne}).

\subsubsection{One}

\begin{figure}[H]
\centering
\includegraphics[scale=0.4]{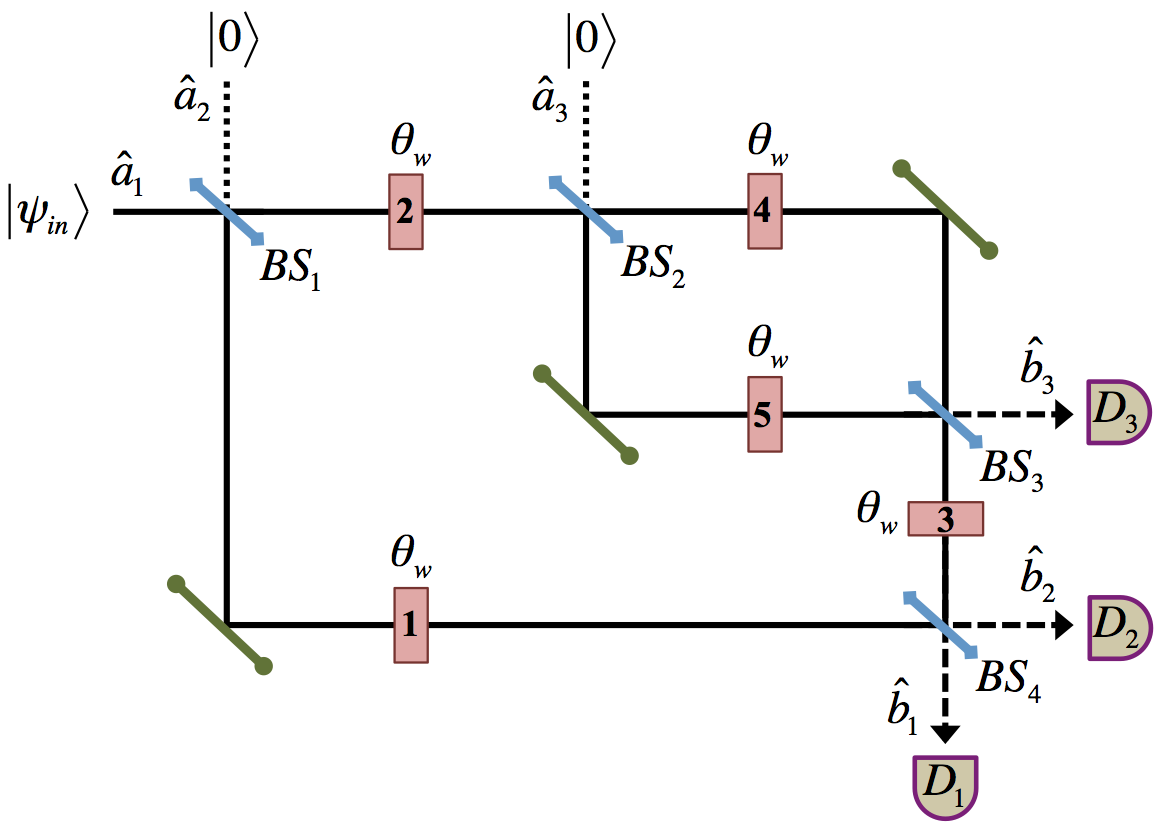}
\caption{(color online) A polarization rotation is added to one, but only one, of the positions, 1 - 5, in the Nested Mach-Zehnder interferometer from Fig. \ref{fig:NestMZI}.}
\label{fig:NestMZIOne}
\end{figure}

In our first scenario, we introduce a rotator in the lower arm of the Nested MZI. See position $(1)$ in Fig. \ref{fig:NestMZIOne}.

The scattering matrix of this device is given by:

\begin{align}
 \mathcal{S}_1 = \begin{pmatrix}
    r_1 r_4 2\overline{\theta}_{w}&  t_1 r_42\overline{\theta}_{w}  & t_4 &  r_1 r_4 \theta_w  & t_1 r_4 \theta_w   & 0 \\
   -r_1 t_4 2\overline{\theta}_{w}  & -t_1 t_4  2\overline{\theta}_{w} & r_4 & - r_1 t_4 \theta_w & -t_1 t_4 \theta_w  & 0 \\
   t_1 & -r_1 & 0 & 0 & 0 & 0 \\
   -r_1 r_4 \theta_w  & -t_1 r_4 \theta_w  & 0  &  r_1 r_4 2\overline{\theta}_{w}  &  t_1 r_4 2\overline{\theta}_{w}  & t_4 \\
   r_1 t_4 \theta_w & t_1 t_4 \theta_w  & 0  &  -r_1 t_4 2\overline{\theta}_{w}  &  -t_1 t_4 2\overline{\theta}_{w}  & r_4 \\
   0 & 0 & 0 & t_1 & -r_1 & 0 
 \end{pmatrix} ,
\end{align}
where $\overline{\theta}_{w} \equiv \sqrt{1-\theta_w^2}/2$.

Assuming the single-photon input state from above, $\ket{\psi_{in}}$, we get the following conditional probabilities for the possible output detections:
\begin{align}
& P(n_{1,H}=1 | \theta_w) =  r_1^2 r_4^2 \big( 1-\theta_w^2 \big), \\
& P(n_{1,V}=1 | \theta_w) =  r_1^2 r_4^2 \theta_w^2 , \\
& P(n_{2,H}=1 | \theta_w) =  r_1^2 t_4^2 \big( 1-\theta_w^2 \big), \\
& P(n_{2,V}=1 | \theta_w) =  r_1^2 t_4^2 \theta_w^2, \\
& P(n_{3,H}=1 | \theta_w) =  t_1^2 , \\
& P(n_{3,V}=1 | \theta_w) =  0 .
\end{align}

The Fisher information of this device is given by:
\begin{equation}
F=  \frac{4}{1-\theta_w^2} r_1^2 .
\end{equation}

Furthermore, the Shannon mutual information is given by:
\begin{equation}
H(\theta_w : M) =  \frac{\ln(108)-4}{3\ln(2)}r_1^2 .
\end{equation}

We see that the information content obtained from the measurement outcomes, as compared to the free-space scenario, is scaled by the square of the reflection coefficient of the first beam-splitter ($BS_1$). The information is reduced exactly by the square of the norm of the wavefunction that is scattered through parts of the interferometer not passing through the polarization rotator in the Schr\"odinger picture.

\subsubsection{Two}
In our next scenario, we consider the rotator to be placed in the upper interferometer path, before the second beam-splitter. The scattering matrix then takes the form of:
\begin{align}
 \mathcal{S}_2 = \begin{pmatrix}
  r_1 r_4 &   t_1 r_4 & t_4 &  0 & 0 & 0 \\
   - r_1 t_4 & - t_1 t_4 & r_4 & 0 & 0 & 0 \\
  t_1  2\overline{\theta}_{w}& -r_12\overline{\theta}_{w} & 0 & t_1 \theta_w & -r_1 \theta_w & 0 \\
   0 & 0 & 0  &  r_1 r_4 &   t_1 r_4 & t_4 \\
  0 & 0 & 0  &  - r_1 t_4 & - t_1 t_4 & r_4 \\
   -t_1 \theta_w & r_1 \theta_w & 0  & t_1 2\overline{\theta}_{w} & -r_1 2\overline{\theta}_{w} & 0
 \end{pmatrix} .
\end{align}

The conditional probabilities of this example are similar to those in Scenario One. However, the dependence on $\theta_w$ is transferred to the third output port, from the first and second in the example above. The conditional probabilities are now given by:
\begin{align}
& P(n^H_1=1 | \theta_w) =  r_1^2 r_4^2, \\
& P(n^V_1=1 | \theta_w) = 0 , \\
& P(n^H_2=1 | \theta_w) =  r_1^2 t_4^2 , \\
& P(n^V_2=1 | \theta_w) = 0 , \\
& P(n^H_3=1 | \theta_w) =  t_1^2 \big( 1-\theta_w^2 \big), \\
& P(n^V_3=1 | \theta_w) =  t_1^2 \theta_w^2.
\end{align}

The Fisher information is given by the expression:
\begin{equation}
\label{Eq:Fish2}
F=  \frac{4}{1-\theta_w^2} t_1^2 .
\end{equation}

Additionally, the Shannon mutual information is scaled similarly, such that it is given by:
\begin{equation}
H(\theta_w : M) =  \frac{\ln(108)-4}{3\ln(2)}t_1^2 .
\end{equation}

The $r_1$ dependency of the previous scenario has, naturally, been transformed into a $t_1$ dependency. Owing to the design of the  NMZI device, the beam-splitters of the inner MZI have no effect on the information of $\theta_w$ if the rotator is placed just after the first beam-splitter.

\subsubsection{Three}
In this scenario we investigate how the above studied properties change if the polarizing rotator is instead placed after the third---but before the fourth---beam-splitter. The scattering matrix is then given by:

\begin{align}
 \mathcal{S}_3 = \begin{pmatrix}
  r_1 r_4 &   t_1 r_4 & t_4 2\overline{\theta}_{w}  &  0 & 0 &  t_4 \theta_w  \\
   - r_1 t_4 & - t_1 t_4 & r_4 2\overline{\theta}_{w} & 0 & 0 &   r_4 \theta_w  \\
   t_1 & -r_1 & 0 & 0 & 0 & 0 \\
   0 & 0 & - t_4 \theta_w   &  r_1 r_4 &   t_1 r_4 & t_4  2\overline{\theta}_{w}\\
  0 & 0 & - r_4 \theta_w  &  - r_1 t_4 & - t_1 t_4 & r_4 2\overline{\theta}_{w}\\
   0 & 0 & 0  & t_1 & -r_1 & 0
 \end{pmatrix} .
\end{align}

The conditional probabilities are given by:
\begin{align}
& P(n^H_1=1 | \theta_w) =  r_1^2 r_4^2 , \\
& P(n^V_1=1 | \theta_w) = 0 , \\
& P(n^H_2=1 | \theta_w) =  r_1^2 t_4^2 , \\
& P(n^V_2=1 | \theta_w) = 0 , \\
& P(n^H_3=1 | \theta_w) =   t_1^2 , \\
& P(n^V_3=1 | \theta_w) =  0 .
\end{align}
The lack of dependence on $\theta_w$ can simply be explained by the fact that the interference effects of the device prohibits any part of the wavepacket to evolve into the spatial location where the rotator is placed in this scenario. 

The Fisher information is given by:
\begin{equation}
F=0 .
\end{equation}
The absence of $\theta_w$ in the conditional probabilities leads to a vanishing Fisher information. The same applies to the Shannon mutual information:
\begin{equation}
H(\theta_w : M) = 0 .
\end{equation}

In the investigation of this scenario, we make the observation of how the introduction of a polarizing rotator, (3) in Fig. \ref{fig:NestMZIOne}, does not alter the output probabilities from the original device in Fig. \ref{fig:NestMZI}. There is never any part of the wavefunction moving from the third to the fourth beam-splitter. Hence, an interaction in this region should not yield any information encoded in the particle.

\subsubsection{Four \& Five}
In Scenario Four and Five we place the polarizing rotator in the upper and lower path of the inner MZI respectively. The scattering matrices of these cases are given by:
\begin{widetext}
\begin{align}
 \mathcal{S}_4 = 
 \begin{pmatrix}
  r_1 r_4 - t_1 t_4  \vartheta_{w}^{-} &     t_1 r_4 + r_1 t_4  \vartheta_{w}^{-} & t_4  \vartheta_{w}^{+} &  t_1 t_4 \overline{\vartheta}_{w} & -r_1 t_4 \overline{\vartheta}_{w} & t_4 \overline{\vartheta}_{w}^2  \\  
    - r_1 t_4 - t_1 r_4   \vartheta_{w}^{-} & -t_1 t_4 +   r_1 r_4   \vartheta_{w}^{-} &  r_4  \vartheta_{w}^{+} &  t_1 r_4 \overline{\vartheta}_{w} & - r_1 r_4 \overline{\vartheta}_{w} &  r_4 \overline{\vartheta}_{w} \\   
    t_1  \vartheta_{w}^{+} & -    r_1  \vartheta_{w}^{+} & -  \vartheta_{w}^{-} &   t_1 \overline{\vartheta}_{w}  & - r_1 \overline{\vartheta}_{w}  &  \overline{\vartheta}_{w} \\
  - t_1 t_4 \overline{\vartheta}_{w} &  r_1 t_4 \overline{\vartheta}_{w} & - t_4 \overline{\vartheta}_{w}    &   r_1 r_4 -  t_1 t_4   \vartheta_{w}^{-} &     t_1 r_4 +  r_1 t_4    \vartheta_{w}^{-} &  t_4   \vartheta_{w}^{+} \\
 - t_1 r_4 \overline{\vartheta}_{w} &  r_1 r_4 \overline{\vartheta}_{w} & - r_4 \overline{\vartheta}_{w} &  - r_1 t_4 -  t_1 r_4  \vartheta_{w}^{-}  & -t_1 t_4 + r_1 r_4   \vartheta_{w}^{-} &  r_4  \vartheta_{w}^{+} \\
- t_1 \overline{\vartheta}_{w}  &  r_1 \overline{\vartheta}_{w}  & - \overline{\vartheta}_{w} &  t_1  \vartheta_{w}^{+} & -   r_1  \vartheta_{w}^{+} & -\vartheta_{w}^{-}  
 \end{pmatrix}  ,
\end{align}
\begin{align}
 \mathcal{S}_5 = 
 \begin{pmatrix}
  r_1 r_4 - t_1 t_4   \vartheta_{w}^{-} &     t_1 r_4 +  r_1 t_4   \vartheta_{w}^{-} & - t_4   \vartheta_{w}^{+} &  - t_1 t_4 \overline{\vartheta}_{w} &  r_1 t_4 \overline{\vartheta}_{w} &  t_4 \overline{\vartheta}_{w}  \\  
    - r_1 t_4 -  t_1 r_4   \vartheta_{w}^{-} & -t_1 t_4 +   r_1 r_4   \vartheta_{w}^{-} & - r_4  \vartheta_{w}^{+} & - t_1 r_4 \overline{\vartheta}_{w} &  r_1 r_4 \overline{\vartheta}_{w} &  r_4 \overline{\vartheta}_{w} \\   
    t_1  \vartheta_{w}^{+} & -    r_1  \vartheta_{w}^{+} &   \vartheta_{w}^{-} & -  t_1 \overline{\vartheta}_{w}  &  r_1 \overline{\vartheta}_{w}  &  \overline{\vartheta}_{w} \\   
  - t_1 t_4 \overline{\vartheta}_{w} & r_1 t_4 \overline{\vartheta}_{w} &  t_4 \overline{\vartheta}_{w}    &   r_1 r_4 -  t_1 t_4  \vartheta_{w}^{+} &     t_1 r_4 +  r_1 t_4    \vartheta_{w}^{+} & - t_4   \vartheta_{w}^{-} \\   
 - t_1 r_4 \overline{\vartheta}_{w} &  r_1 r_4 \overline{\vartheta}_{w} &  r_4 \overline{\vartheta}_{w} &  - r_1 t_4 -  t_1 r_4  \vartheta_{w}^{+}  & -t_1 t_4 + r_1 r_4   \vartheta_{w}^{+} & - r_4  \vartheta_{w}^{-} \\  
- t_1 \overline{\vartheta}_{w}  &  r_1 \overline{\vartheta}_{w}  &  \overline{\vartheta}_{w} &   t_1  \vartheta_{w}^{-} & - r_1  \vartheta_{w}^{-} &   \vartheta_{w}^{+}  
 \end{pmatrix}  .
\end{align}
\end{widetext}
where $\vartheta_{w}^{\pm} \equiv (1 \pm \vartheta_{w})/2$ and where we temporarily make a superficial change of variables such that $\vartheta_w \equiv \sqrt{1-\theta_w^2}$.

For the two scenarios of introducing the polarization rotator inside the nested part of the interferometer, the conditional probabilities take more complicated forms:
\begin{align}
& P(n^H_1=1 | \vartheta_w) =  \frac{1}{4}(2r_1 r_4 -t_1 t_4 (1-\vartheta_w))^2 , \\
& P(n^V_1=1 | \vartheta_w) = \frac{1}{4} t_1^2 t_4^2(1-\vartheta_w^2), \\
& P(n^H_2=1 | \vartheta_w) =  \frac{1}{4}(2r_1 t_4 + t_1 r_4 (1-\vartheta_w))^2 , \\
& P(n^V_2=1 | \vartheta_w) = \frac{1}{4} t_1^2 r_4^2 (1-\vartheta_w^2) , \\
& P(n^H_3=1 | \vartheta_w) =   \frac{1}{4}t_1^2 (1+\vartheta_w)^2, \\
& P(n^V_3=1 | \vartheta_w) =   \frac{1}{4}t_1^2 (1-\vartheta_w^2) .
\end{align}

The corresponding Fisher information of the device in Fig. \ref{fig:NestMZIOne}, with a polarization rotator in the nested part ($(4)$ or $(5)$), is given by:
\begin{equation}
\label{Eq:NestFish}
F(\theta_w) =  \frac{2}{1-\theta_w^2} t_1^2 .
\end{equation}

In accordance with Eq. \ref{Eq:OptFish}, we see that the Fisher information is proportional to how much of the wavepacket that---in the Schr\"odinger picture---has passed through the rotator in the device. In the scenarios of this subsection, the part of the wavefunction that travelled through the rotator in Scenario Two, is halved by the second beam-splitter before it is allowed to interact with the rotator. Hence,  $F$ (in Eq. \ref{Eq:NestFish}) is halved as compared to its value in Scenario Two (Eq. \ref{Eq:Fish2}).

\begin{table}[b]
\caption{\label{table:Pade}%
Numerical constants in $[6/4]$ Padé approximation of Eq. \ref{Eq:Mutual}.
}
\begin{ruledtabular}
\begin{tabular}{ccc}
\textrm{$i$} &
\textrm{$a_i$} &
\textrm{$b_i$}\\
\colrule
$2$ & $\frac{-3+\ln{(2)} + 3\ln{(3)}}{3\ln{(2)}}$ & $-1$ \\
$4$ & $\frac{25-6\ln{(2)} + 25\ln{(3)}}{18\ln{(2)}}$ & $\frac{-1}{10(-7+3\ln{(3)})}$
 \\
$6$ & $\frac{254 - 3\ln{(2)} -429\ln{(3)} + 180 \ln{(3)}^2}{90(-7+3\ln{(3)})\ln{(3)}}$ &  
\end{tabular}
\end{ruledtabular}
\end{table}

\begin{figure}
\centering
\includegraphics[scale=0.32]{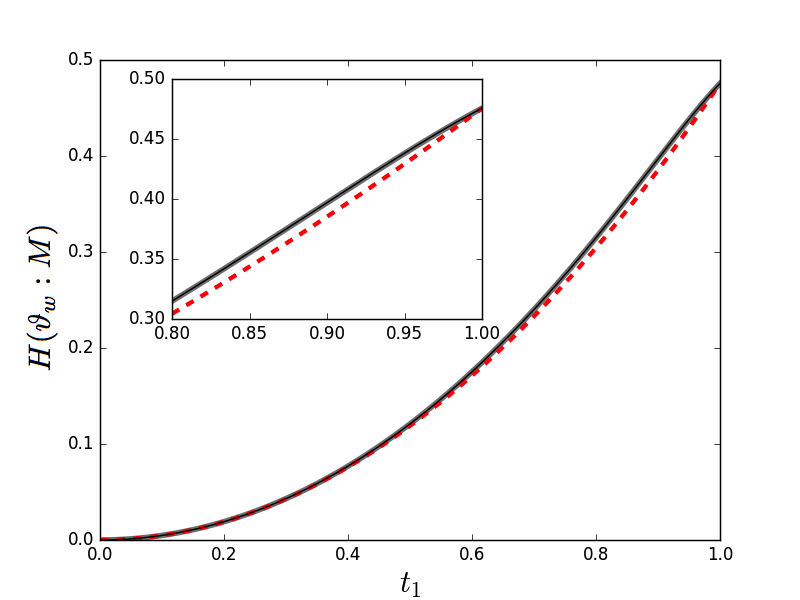}
\caption{(color online) The mutual information between the polarization rotation, $\vartheta_{w}$, and the measurement outcomes, $\{ M_i\}$, as a function of beam-splitter transmission, $t_1$, as described in the text. The solid black line shows the true curve of Eq. \ref{Eq:Mutual}, the thick grey line shows the Padé approximation (virtually indistinguishable from the true curve) and the red dashed line shows the second order Taylor expansion.}
\label{fig:ShannonCurves}
\end{figure}

\drma{Whilst the Fisher information preserves the simple form of Eq. \ref{Eq:OptFish}}, the corresponding Shannon mutual information for these devices is complicated and not very informative. However, by making the assumption that $t_4=r_1$ and $r_4=t_1$, we can simplify the expression of the mutual information such that:

\begin{gather}
H(\vartheta_{w} : M) =  \frac{1}{ 3 \ln{(2)} t_1^2} \bigg[   -2r_{1}^3 \ln{(r_1^2)}  \nonumber  \\  +t_1^2  \Big( 3\ln{(3)}+t_1^2 \big(\ln{(2)}-1   \big)    \nonumber \\ -2  -\big( 3r_1^2 + t_1^4 \big)\ln{\big( 3r_1^2 + t_1^4 \big)} \Big) \bigg].
\label{Eq:Mutual}
\end{gather}
Furthermore, we can approximate this expression. For scenarios where $t_1\approx 0$, the Shannon mutual information is successfully modelled by a second order term:
\begin{equation}
H(\vartheta_{w} : M) \approx \frac{-3+\ln{(2)+3\ln{(3)}}}{3\ln{(2)}} t_1^2 .
\end{equation}
To obtain an even better model \drma{(for $0 \leq t_1 \leq 1$)} we can use a Padé approximant \cite{bPress92} of order $[6/4]$:
\begin{equation}
H(\vartheta_{w} : M) \approx \frac{a_2 t_1^2 + a_4 t_1^4 +a_6  t_1^6}{1+b_2 t_1^2 + b_4 t_1^4} ,
\end{equation}
with constants $a_i$ and $b_j$ for $i \in \{2,4,6\}$ and $j \in \{2,4\}$ given in Table \ref{table:Pade}. Fig. \ref{fig:ShannonCurves} shows the mutual information from Eq. \ref{Eq:Mutual} and the two approximations as functions of $t_1$. The second order Taylor expansion (for $t_1 \leq 0.4$) and the full Padé approximation, model the true curve within mean squared errors of $3.1 \times 10^{-8}$ and $2.8 \times 10^{-9}$ respectively.

\drma{After having established the bounds on parameter estimation with NMZI structures, we proceed to evaluate the counterfactuality of the Type \rom{1} protocols, which are based on such devices.}

\section{Evaluation of Type \rom{1} Counterfactual Communication} \label{Sec:Type1Vio}

Fig. \ref{fig:ChainedNMZI} shows a chained nested Mach-Zehnder interferometer.  The first proposal of direct CFC \cite{Salih13} is based on such a device. The chained NMZI is divided such that the top right part of the individual NMZIs are in Bob's laboratory. The communication scheme allows Alice to input a photon in the top left path (solid black line). If Bob wishes to transmit a logical $0$, he leaves all paths open. If he, instead, wishes to transmit a logical $1$, he blocks all paths with his detectors, $D_B$. Bob's action, together with the number of inner and outer beam-splitters ($M$ and $N$ respectively), sets the output statistics in Alice's laboratory (detections at $D_1$ or $D_2$). The outer and inner beam-splitters have their transmission coefficients set such that $t_{n=1,\ldots,N} = \sin{(\pi / 2N)}$ and $t_{m=1,\ldots,M} = \sin{(\pi / 2M)}$ respectively. In theory, for an infinitely large number of beam-splitters, the photon \drma{can be made to} end up at $D_2$ \drma{with $P_{D_2}=1$} (logical $1$), or $D_1$ \drma{with $P_{D_1}=1$} (logical $0$), if Bob inserts or does not insert $D_B$ in his laboratory, respectively \cite{Salih13}.

\begin{figure}
\centering
\includegraphics[scale=0.2]{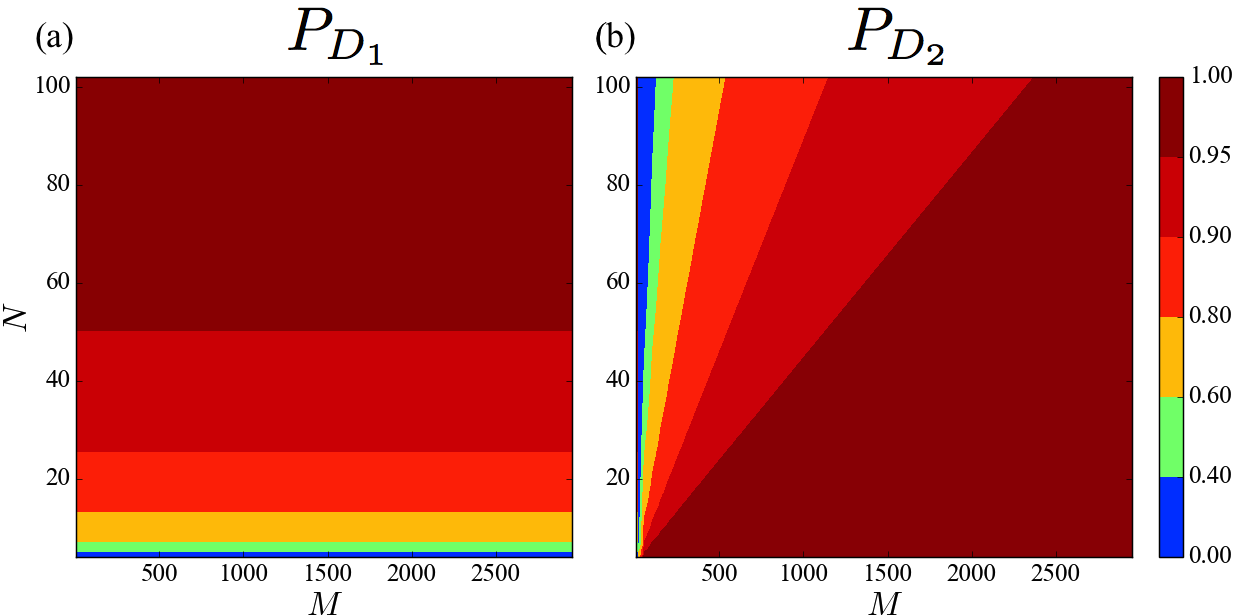}
\caption{(color online) Probability of detection at $D_1$ (a) and $D_2$ (b) if Bob unblocks and blocks his path in Fig. \ref{fig:ChainedNMZI} respectively.}
\label{fig:ChainedNMZIProbs}
\end{figure}

Ref. \cite{Salih13} assumes that the evolution of the interrogating particle in the above described scheme can be modeled by perfectly unitary rotation matrices. For reference, we numerically calculate the detection probabilities of $D_1$ and $D_2$ detections in the scenarios of Bob keeping his laboratory open and blocked respectively. These probabilities are shown in Fig. \ref{fig:ChainedNMZIProbs} and are in accordance with those calculated in Ref. \cite{Salih13}. 

Fig. \ref{fig:ChainedNMZIProbs} suggests that for a communication scheme with low logical bit errors we need $M \gg N$. \drma{We see that success rates of $95 \%$ for the logical $0$ (Fig. \ref{fig:ChainedNMZIProbs}(a))  requires large $N \approx 50$. For such values of $N$ we need $M \approx 1200$ to keep the same success rates for the logical $1$ (Fig. \ref{fig:ChainedNMZIProbs}(b)).  Hence,} as discussed in Ref. \cite{ArvShukur16},  a success rate of about $95 \%$ requires a total of approximately $60000$ beam-splitters to be used.

\subsection{Single NMZI}

Ref. \cite{Vaidman14, Salih14, Vaidman15} suggests that the conceptual problem of the chained NMZI in Ref. \cite{Salih13} can be reduced to a study of a single NMZI device by considering pre- and post-selected events. \color{black}Following this reduction, the mentioned references analytically analyse Type \rom{1} ``counterfactual" schemes based on single NMZI structures. However, to our knowledge there exist no rigorous proof that this reduction is an adequate representation of Salih's scheme. Crucially, the single evaluation of the protocol that they consider does not allow for the transmission of logical bits. In this section we take an alternative approach. 

\subsubsection{Analytical Analysis}

\begin{figure}
\centering
\includegraphics[scale=0.41]{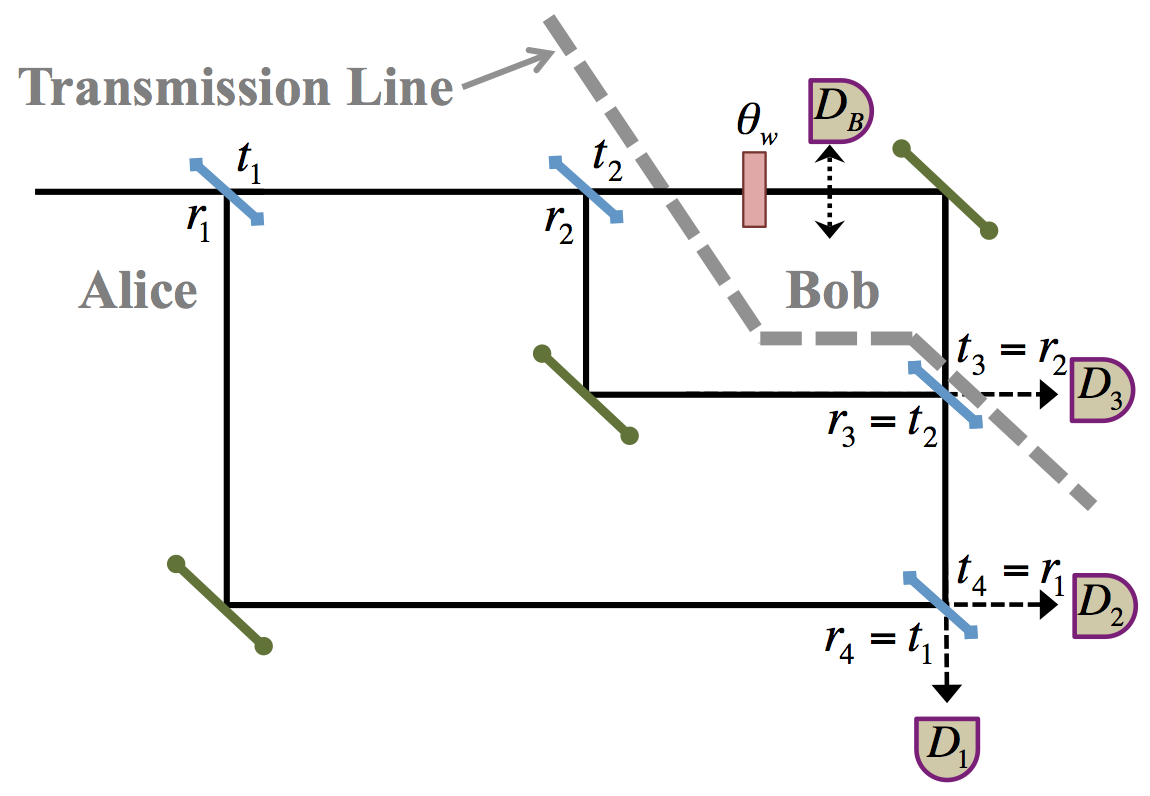}
\caption{(color online) The Nested Mach-Zehnder interferometer as used in the communication scheme presented in the text. }
\label{fig:NMZIComm}
\end{figure}

Instead of treating the single NMZI device as a representation of Salih's scheme, we evaluate a post-selected Type \rom{1} protocol that actually allows for communication. Consider Fig. \ref{fig:NMZIComm}. We pre-select our states such that the input is the usual $\ket{\psi_{in}}=a^{\dagger}_{1,H}\ket{0}$ from before. Bob has the option of introducing some absorbing object in his laboratory. Furthermore, we include a weak polarization rotation in Bob's laboratory to mimic some disturbance in the device. We also post-select our states such that we exclude the events of absorption in Bob's laboratory, by $D_B$ or $D_3$.

If Bob wishes to transmit a logical $1$ to Alice, he introduces the absorbing object in his laboratory. The renormalized output probabilities are then given by:

\begin{align}
& P_1(n^H_1=1 | n^{H,V}_{3,B}=0) =  \mathcal{N}_1 \Big( r_1 r_4 - r_2 t_1 t_2 t_4        \Big)^2, \\
& P_1(n^V_1=1 |n^{H,V}_{3,B}=0) = 0 , \\
& P_1(n^H_2=1 | n^{H,V}_{3,B}=0) =  \mathcal{N}_1 \Big( r_1 t_4 + r_2 r_4 t_1 t_2 \Big)^2, \\
& P_1(n^V_2=1 | n^{H,V}_{3,B}=0) = 0 ,
\end{align}
with $\mathcal{N}_1 \equiv  \Big( 1 -  t_1^2 \big( r_2^4+t_2^2 \big)   \Big)^{-1} $.

We set the beam-splitter parameters according to Fig. \ref{fig:NMZIComm}, with $r_2=t_2=1/\sqrt{2}$. This simplifies our expression such that:
\begin{align}
& P_1(n^H_1=1 | n^{H,V}_{3,B}=0) =  \frac{r_1^2 t_1^2}{3r_1^2+1}, \\
& P_1(n^V_1=1 | n^{H,V}_{3,B}=0) = 0 , \\
& P_1(n^H_2=1 | n^{H,V}_{3,B}=0) =  \frac{\big( r_1^2 +1\big)^2}{3r_1^2+1}, \\
& P_1(n^V_2=1 | n^{H,V}_{3,B}=0) = 0 .
\end{align}
The corresponding value for the conditioned Fisher information of this $1$-bit protocol is:
\begin{equation}
F^1=0 .
\end{equation} 
This is expected, as any part of the wavefunction that interacts with the rotator is then absorbed by $D_B$ and can thus not reach Alice.

Now, consider the process in which Bob instead wishes to transmit a logical $0$ to Alice. He then leaves his laboratory free, without any absorbing objects. The renormalized probabilities of detecting the particle in the following states are given by:

\begin{align}
 P_0(n^H_1=1 | \theta_w, n^{H,V}_{3,B}=0) = & \mathcal{N}_0 \Big( r_1 r_4 + r_2 t_1 t_2 t_4 \nonumber \\ & \times \big( 2 \overline{\theta}_{w} - 1 \big)       \Big)^2, \\
 P_0(n^V_1=1 | \theta_w, n^{H,V}_{3,B}=0) = & \mathcal{N}_0 \Big(  r_2^2 t_1^2 t_2^2 t_4^2 \theta_w^2  \Big) , \\
 P_0(n^H_2=1 | \theta_w, n^{H,V}_{3,B}=0) = &  \mathcal{N}_0 \Big( r_1 t_4 \nonumber - r_2 r_4 t_1 t_2  \\ & \times \big( 2 \overline{\theta}_{w} - 1 \big)       \Big)^2, \\
 P_0(n^V_2=1 | \theta_w, n^{H,V}_{3,B}=0) = & \mathcal{N}_0 \Big(  r_2^2 r_4^2 t_1^2 t_2^2  \theta_w^2  \Big)  ,
\end{align}
 with $\mathcal{N}_0 \equiv  \Big( 1- t_1^2 \big(r_2^4+t_2^4+4 r_2^2 t_2^2 \overline{\theta}_{w} \big) \Big)^{-1}$. Again, we set the beam-splitter parameters according to Fig. \ref{fig:NMZIComm}, with $r_2=t_2=1/\sqrt{2}$. The probabilities then simplify to:
\begin{align}
& P_0(n^H_1=1 | \theta_w, n^{H,V}_{3,B}=0) = \mathcal{N}_0^{\prime}  r_1^2 t_1^2 \big( 1 + 2 \overline{\theta}_{w} \big)^2  , \\
& P_0(n^V_1=1 | \theta_w, n^{H,V}_{3,B}=0)  = \mathcal{N}_0^{\prime} r_1^2 t_1^2\theta_w^2   , \\
& P_0(n^H_2=1 | \theta_w, n^{H,V}_{3,B}=0)  = \mathcal{N}_0^{\prime}  \big(1+ r_1^2 -  2 t_1^2 \overline{\theta}_{w} \big)^2 , \\
& P_0(n^V_2=1 | \theta_w, n^{H,V}_{3,B}=0)  =   \mathcal{N}_0^{\prime} t_1^4\theta_w^2  ,
\end{align}
 where $ \mathcal{N}_0^{\prime} \equiv \Big(4-2t_1^2 \big( 1 + 2 \overline{\theta}_{w} \big)\Big)^{-1}$.

 For the $0$-bit scheme considered above, we then obtain:
\begin{equation}
F^{0}=\frac{t_1^2}{1-\theta_w^2} .
\end{equation}
This can be compared to the $1$-bit scheme, where the value is $F^{1}=0$, such that there is no information about the angle $\theta_w$ given to Alice.

The two processes described in this section can be used in order to transmit information from Bob to Alice in a scenario where Bob only has access to the inner part of the NMZI (see Fig. \ref{fig:NMZIComm}). However, in order for our setup to be representative of the behaviour of the chained NMZI structure from the section above, we set $\theta_w \ll t_1 \ll r_1$. Hence, the polarization rotation will have a minute impact on the probability outputs. For $t_1 \ll r_1$, the probability distributions for the $0$- and $1$-bit processes are very similar. We see that in both processes of the scheme, Alice will detect the state $\ket{n^H_2=1}$ with high probability. Thus, in order for Alice to obtain Bob's choice of bit-value, with high probability, each logical bit has to be decoded from a larger bit string.

The communication scheme is as follows: Alice sends a number, $n_\gamma$, of single photons (excluding the particles that do not fulfil the post-selection criterion and are absorbed by either $D_B$ or $D_3$)  into the device of Fig. \ref{fig:NMZIComm}, one after another. Depending on what logical bit Bob wishes to transmit, he either inserts detector $D_B$ or leaves his laboratory open, for all the $n_\gamma$ particles. Alice makes subsequent particle detections of the $n_\gamma$ events: $\ket{n^H_1=1}$, $\ket{n^H_2=1}$, $\ket{n^V_1=1}$ or $\ket{n^V_2=1}$. If she measures any event in $\ket{n^V_1=1}$ or $\ket{n^V_2=1}$, she knows with certainty that a logical $0$ was sent. However, owing to the fact that $\theta_w$ is very small, the accumulative probability of these events is also small. Hence, Alice will, with high probability, have to use the statistics of $\ket{n^H_1=1}$ and $\ket{n^H_2=1}$ detections to infer the logical bit.   From the number of particles, $q$, that Alice measure in the $\ket{n^H_1=1}$ state, she decides whether Bob sent a logical $0$ or a logical $1$. 

The question of interest is: what number, $n_\gamma$, of single-photon evaluations of the device, allows for an effective communication scheme with a limited number of errors?

We re-define $P_1 \equiv P_1(n^H_1=1 | n^{H,V}_{3,B}=0)$ and $ P_0 \equiv P_0(n^H_1=1 | \theta_w, n^{H,V}_{3,B}=0)$. For small $t_1$ and $\theta_w \ll t_1$, we see that $P_1 < P_0$. Alice will thus note down a $1$ every time $q < q^\prime$. In the limit of long message strings, Bob will produce logical $0$s and $1$s at the same rates, and the exact value of $q^\prime$ is given by:
\begin{equation}
q^\prime = \floor*{ \frac{n_\gamma \ln{\Big( \frac{1-P_1}{1-P_0} \Big)} }{\ln \Big( \frac{P_0}{P_1} \Big)-\ln\Big( \frac{1-P_0}{1-P_1} \Big)}} .
\end{equation}
The probability for a non-faulty logical bit-transmission is thus:
\begin{align}
P_{succ.}  & =  \frac{1}{2} \sum_{q=0}^{q^\prime} \frac{n_\gamma! P_1^q (1-P_1)^{n_\gamma-q}}{q!(n_\gamma-q)!} \nonumber \\ & + \frac{1}{2} \sum_{q=q^\prime + 1}^{n_\gamma} \frac{n_\gamma! P_0^q (1-P_0)^{n_\gamma-q}}{q!(n_\gamma-q)!} .
\label{eq:kProb}
\end{align}

Eq. \ref{eq:kProb} can be used to numerically find an acceptable value of $n_\gamma$, given the parameters of the setup. However, in order to evaluate the setup discussed in Ref. \cite{ Vaidman13, Vaidman13-2, Salih13, Vaidman14, Salih14, Vaidman16}, we need the transmission coefficient to be small ($t_1 =\sin{(\pi / 2 N)} \ll 1$,  with $N \gg 1$). As $t_1$ is small, we can, by the central limit theorem, assume that $n_\gamma$ has to be large and that the two bit-processes will generate normally distributed events. The two processes will each have a mean situated at $P_1$ and $P_0$ respectively. Their respective standard deviations will be given by:
\begin{equation}
\sigma_{{i}} = \sqrt{\frac{P_{i} (1-P_{i})}{n_\gamma}} ,
\end{equation}
where $i=0,1$, which decreases reciprocally with the square-root of $n_\gamma$. For Alice to be able to distinguish between the logical $0$ and $1$ bits correctly with probability $1-\epsilon$, we require that:
\begin{equation}
n_\gamma \geq \Bigg( \Phi^{-1}(\epsilon) \frac{\sqrt{P_0(1-P_0)} + \sqrt{ P_1(1-P_1) }}{P_0-P_1} \Bigg)^2 ,
\end{equation}
where  $\Phi^{-1}(\epsilon)$ is the inverse of the standard normal cumulative distribution function. 

We can Taylor expand $n_{\gamma}$ for small values of $t_1$ such that:
\begin{equation}
n_\gamma \geq  \big( \Phi^{-1}(\epsilon) \big)^2  \frac{4}{t_1^2} + \mathcal{O}(t_1^{-1}) .
\end{equation}

As the Fisher information scales linearly with the number of evaluations of the channel, $n_\gamma$, our counterfactual violation strength for a Type \rom{1} logical $0$-bit is given by:
\begin{equation}
\mathcal{D} = n_\gamma  \frac{F^{0}}{F_{free}} \gtrsim \big( \Phi^{-1}(\epsilon) \big)^2 .
\end{equation}
For a success rate of roughly $95 \%$, we thus obtain a value of $\mathcal{D} \approx 2.7$, and we conclude that the CFC scheme described in this subsection is no more counterfactual than a free space evolution of particles between Alice and Bob. 

\subsubsection{Simulation of Quantum Evolution}

\begin{figure}
\centering
\includegraphics[scale=0.18]{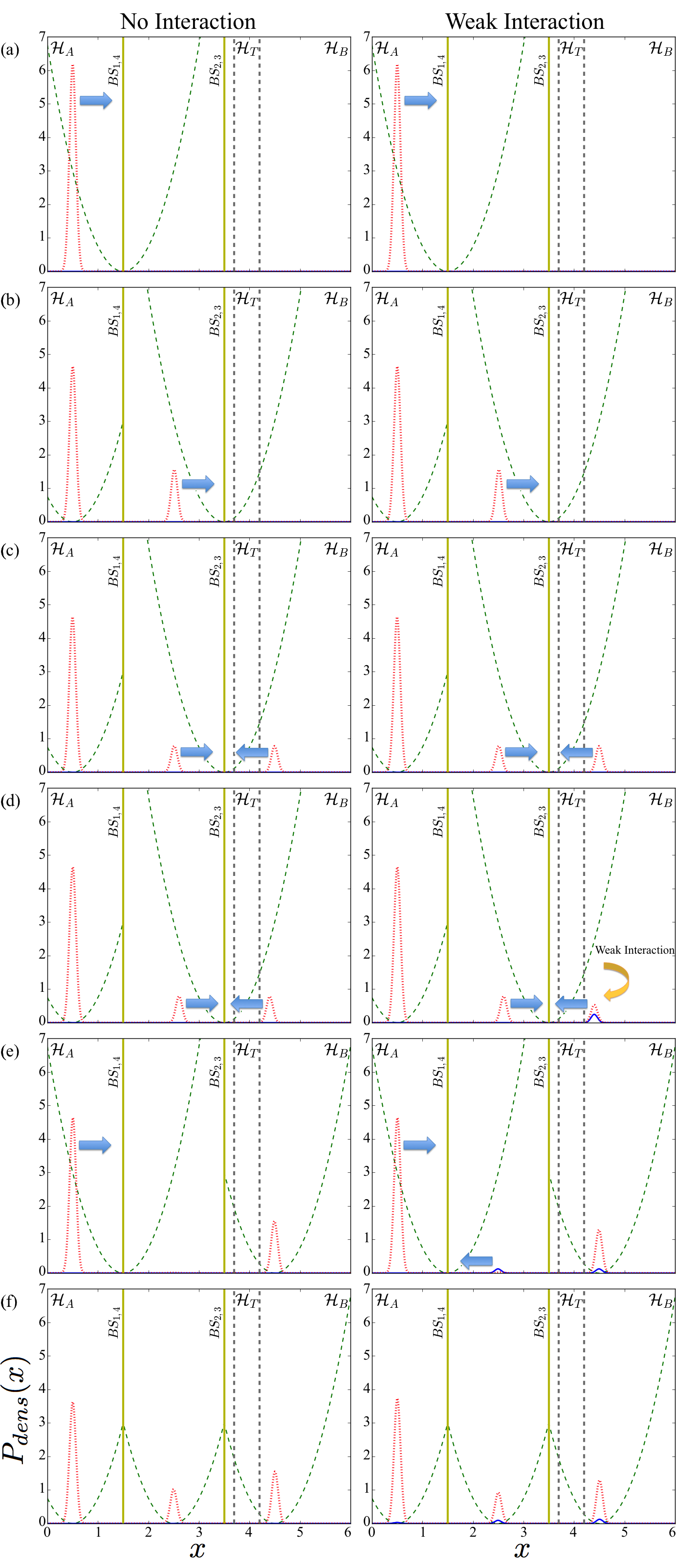}
\caption{(color online) This figure shows the quantum evolution (time steps (a) to (f)) of the probability density distribution of a spin-$\frac{1}{2}$ particle in a nested Mach-Zehnder interferometer with and without a weak spin-rotation interaction (exaggerated for visibility) in Bob's laboratory. The dotted red and solid blue curves indicate spin up and spin down components of the wavefunction respectively. The dashed green curves show the potentials. Beam-splitters are denoted with vertical yellow lines. The spatial components are indicated with the vertical dashed grey lines.  }
\label{fig:SimNMZI}
\end{figure}

In order to illustrate the origin of a counterfactual violation inside a NMZI device, we provide a numerical simulation of the time-dependent Schr\"odinger equation. We simulate a massive Gaussian spin-$\frac{1}{2}$ particle that propagates through a NMZI, which we have mapped onto a linear 1D structure. The Hamiltonian to implement such an evolution can be tailored as in Ref. \cite{ArvShukur16}. \drma{This allows us to design a toy model for the wavefunction evolution in a NMZI.} The solution is calculated by a GPU-boosted version of the Staggered Leapfrog algorithm as in Refs. \cite{Owen14, ArvShukur17}. Fig. \ref{fig:SimNMZI} shows the evolution of the wavefunction. 

The Hamiltonian has been tailored such that the beam-splitter parameters are given by: $t_2=r_2=t_3=r_3 = \frac{1}{\sqrt{2}}$, $t_1=t_4=\frac{1}{2}$ and $r_1=r_4=\frac{\sqrt{3}}{2}$.

As can be seen from Fig. \ref{fig:SimNMZI}, the effect of the weak interaction in Bob's laboratory (right frame in Fig. \ref{fig:SimNMZI}(d)) is to distort the interaction on the beam-splitter (between Fig. \ref{fig:SimNMZI}(d) and \ref{fig:SimNMZI}(f)). In the scenario of no interaction, the second passage through the right beam-splitter causes the middle and right parts of the wavepacket to interfere constructively and propagate to Bob's laboratory, never to return to Alice. However, a weak interaction in Bob's laboratory allows for a fraction of the wavepacket (right frame in Fig. \ref{fig:SimNMZI}(e)) to propagate back towards Alice's laboratory and interfere on the left beam-splitter. Hence, the probability density distributions around $x=0.5$ and $x=2.5$ in Alice's laboratory are different depending on whether or not a weak interaction took place. This is why  the Type \rom{1} counterfactuality is satisfied, only if absolutely pure quantum channels are present in the NMZI device.

\subsection{Chained NMZIs}

\begin{figure}
\centering
\includegraphics[scale=0.28]{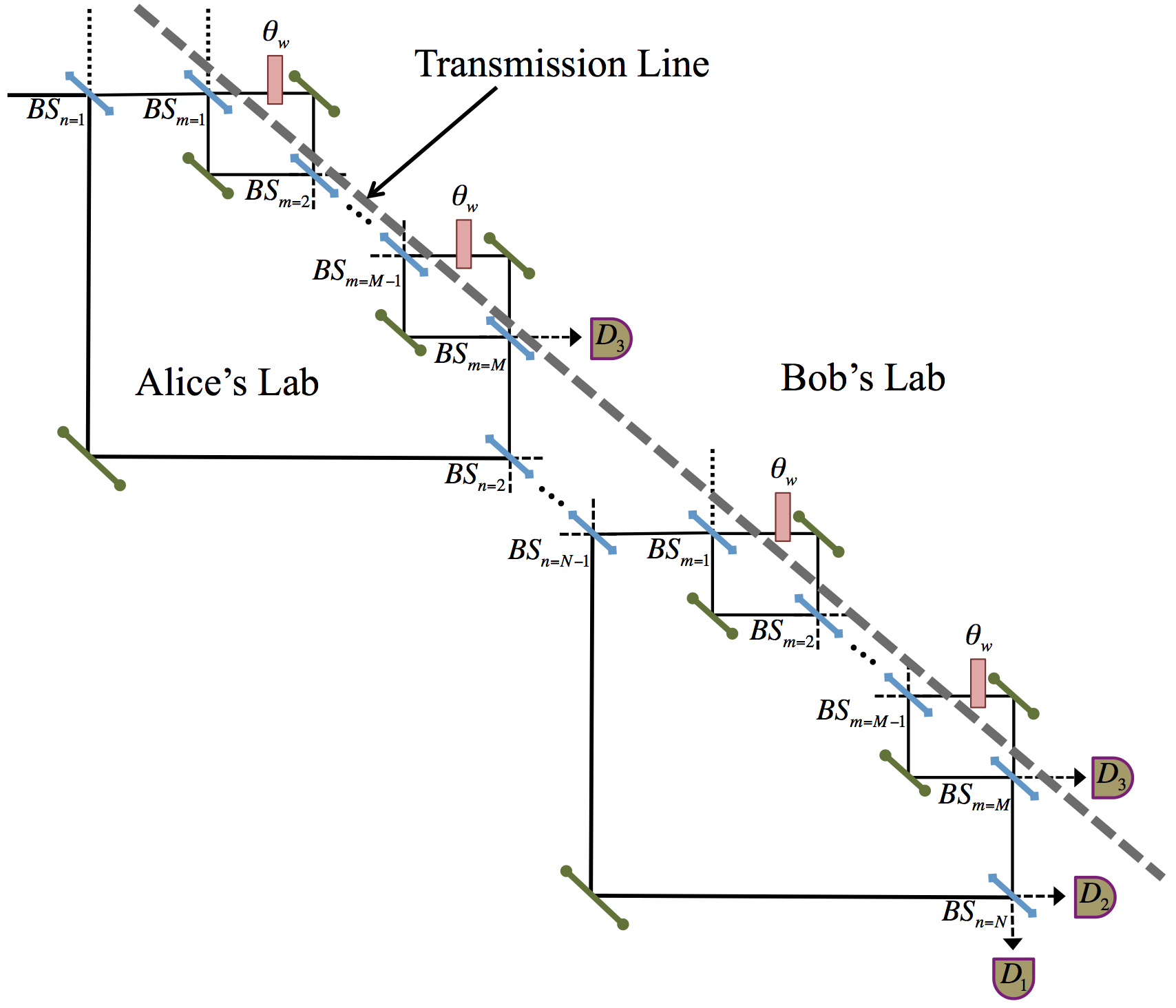}
\caption{(color online) The logical $0$ process for the chained nested Mach-Zehnder interferometer suggested for CFC in Ref. \cite{Salih13}. The weak polarization interactions mimic realistic systematic errors in the quantum channels of Bob's laboratory.}
\label{fig:ChainedNMZI_Weak}
\end{figure}

\drma{In the previous subsection, we analysed a reduced NMZI CFC scheme. However, an extension of the analysis to the originally proposed chained NMZI protocol of Ref. \cite{Salih13} is not straightforward. (Previous attempts have been heavily criticised \cite{Vaidman13, Li13, Vaidman13-2, Salih13, Vaidman14, Salih14, Li15, Vaidman15, Vaidman16, Li16, Griffiths16}). In general, complications arise from the multiple paths ($(N-1) \times (M-1)$ in Fig. \ref{fig:ChainedNMZI}) in and out of Bob's laboratory. In terms of our approach, this complicates the concept of ``presence". Nevertheless, the Fisher information with respect to the weak disturbance in Bob's laboratory can still be calculated---and generally have significantly larger values than those calculated with Eq. \ref{Eq:OptFish}. An analytical analysis of the Fisher information in the chained NMZI devices yields a complicated non-informative expression, even for small numbers of $M$ and $N$. Instead, a numerical finite difference method allows for a comprehensive approximation of the Fisher information. This allows us to calculate the counterfactual violation strength (Eq. \ref{Eq:CountVio}).} 

\begin{figure}
\centering
\includegraphics[scale=0.35]{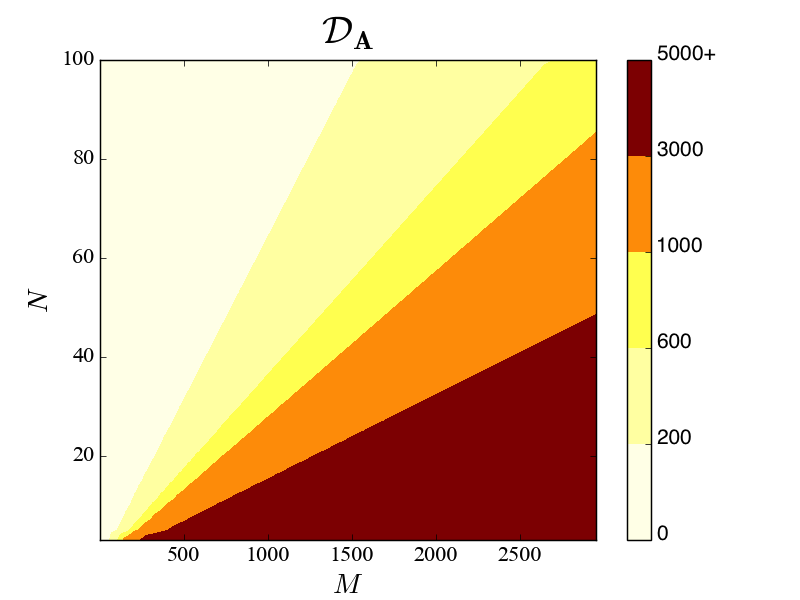}
\caption{(color online) The spatially conditioned Type \rom{1} counterfactual violation strength as a function of the beam-splitter numbers $N$ and $M$ for the scenario of Bob not introducing his detectors in Fig. \ref{fig:ChainedNMZI} or \ref{fig:ChainedNMZI_Weak}.}
\label{fig:ChainedNMZICountVioStrength}
\end{figure}

We now calculate the quantum evolution of the Type \rom{1} logical $0$ process (no $D_B$), with a polarization rotation of $\theta_w$ in every inner MZI in Bob's laboratory. (See Fig. \ref{fig:ChainedNMZI_Weak}). In accordance with the previous sections of this paper, the weak rotations mimic disturbances of realistic quantum channels. Fig. \ref{fig:ChainedNMZICountVioStrength} shows the spatially conditioned (i.e. $F \rightarrow F_{\bm{A}}$ in Eq. \ref{Eq:CountVio}) Type \rom{1} counterfactual violation strength, $\mathcal{D}_{\bm{A}}$, as a function of $N$ and $M$, assuming that Bob leaves his path open and that the polarization rotations are carried out with a weak polarization parameter: $\theta_w = 10^{-6} \ll M^{-1} $. A simple calculation shows that $\mathcal{D} \geq \mathcal{D}_{\bm{A}}$, such that Fig. \ref{fig:ChainedNMZICountVioStrength} can be used as a lower bound on the Type \rom{1} counterfactual violation strength in the device of Fig. \ref{fig:ChainedNMZI_Weak}.

As stated before, large values of $M$ and $N$ are needed to carry out the direct communication scheme with high success probability. For such values, the counterfactual violation strength of the chained NMZI communication scheme is many orders of magnitude larger than unity. For realistic quantum channels, we can thus, based on our counterfactuality measure, conclude that the suggested communication scheme of Ref. \cite{Salih13} is, \textit{de facto}, not counterfactual.

\section{Evaluation of Type \rom{2} Counterfactual Communication}
\label{Sec:Type2Vio}
We now consider the Type \rom{2} protocol suggested by Arvidsson-Shukur and Barnes \cite{ArvShukur16}, which relies on a chained MZI (CMZI) device. Such a device is shown in Fig. \ref{fig:CMZI}.

In this protocol, Alice sends a single-photon state into the upper left input path of the device (solid black line). The photon then enters the transmission line, which shares a CMZI device with Bob's laboratory. Bob has the possibility of inputting detectors in his path or leaving it open. If Bob leaves his path open, the quantum evolution of the photon in the device will lead it to emerge onto detector $D_2$ in Bob's laboratory, without the wavepacket ever spreading into Alice's laboratory after it first left it. However, if Bob instead inputs detectors, $D_B$, after each beam-splitter, the wavepacket will either be absorbed by one of them or collapse onto a state in the lower part of the CMZI device that can re-enter Alice's laboratory to be detected by detector $D_1$. In the limit of large $N$ and inserted detectors, the quantum Zeno effect can make the probability of re-entering in Alice's laboratory arbitrarily close to unity. Hence, Bob's action of either leaving his path free or inserting detectors affects the detection probabilities in Alice's laboratory and---in the limit of large $N$---allows her to deduce Bob's action with high probability of success.\footnote{The protocol presented in Ref. \cite{ArvShukur16} shows how a clever logical bit-encoding scheme can take the probability of success close to unity, even for imperfect beam-splitters and $N \leq 7$.}

This protocol is conceptually different from the one presented in Ref. \cite{Salih13}. That protocol suggests a scenario where the photon would never travel from Bob to Alice or vice versa. The protocol described in Ref. \citep{ArvShukur16} indeed never sees the photon wavefunction travel from Bob to Alice. It does, however, allow for the wavefunction to propagate from Alice to Bob. The protocol is, nevertheless, counterfactual according to the Type \rom{2} definition, which allows particles to travel in the opposite direction to the message.  

\begin{figure}
\centering
\includegraphics[scale=0.32]{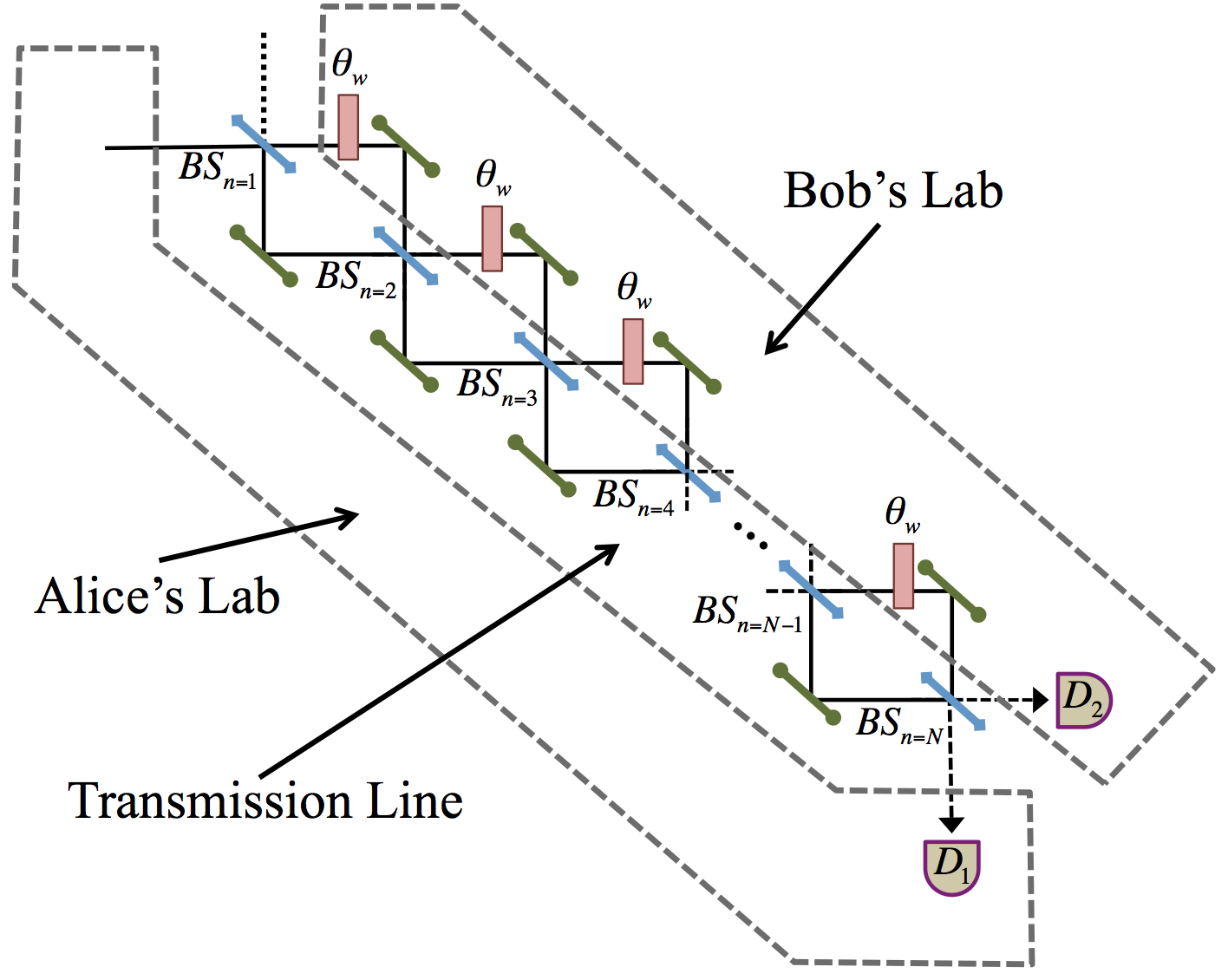}
\caption{(color online) The logical $0$ process of the chained Mach-Zehnder interferometer CFC scheme of Ref. \cite{ArvShukur16}. Again, the weak polarization interactions mimic realistic systematic errors in the quantum channels of Bob's laboratory. }
\label{fig:CMZI_Weak}
\end{figure}

\begin{figure}
\centering
\includegraphics[scale=0.28]{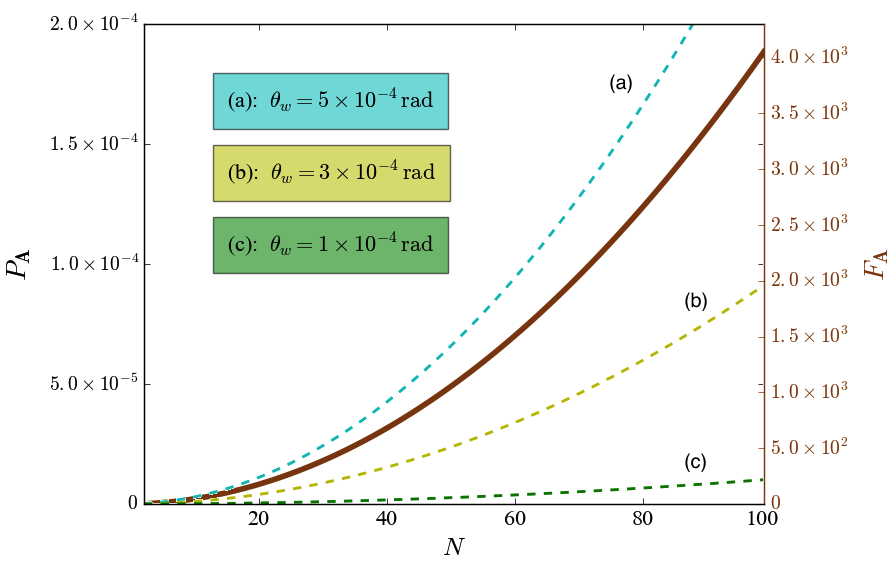}
\caption{(color online) The Type \rom{2} logical $0$ (see Ref. \cite{ArvShukur16}) counterfactual violation strength, $P_{\bm{A}}$ (dashed lines read on the left $y$-axis), and the spatially conditioned Fisher information, $F_{\bm{A}}$ (solid line read on the right $y$-axis), as functions of the beam-splitter number, $N$.  }
\label{fig:CMZIStrength}
\end{figure}

We simulate the Type \rom{2} counterfactual violation strength ($P_{\bm{A}}$ from Eq. \ref{Eq:CountVioProb}) per photon transport through the CMZI with Bob's path open, as a function of $N$. The weak polarization rotators are again inserted in each separate MZI of Bob's laboratory. (See Fig. \ref{fig:CMZI_Weak}). $P_{\bm{A}}$ is dependent of $\theta_w$ and we present results  for various values of $\theta_w$ in Fig. \ref{fig:CMZIStrength}.

The CMZI CFC protocol can be carried out with a high success rate with less than $100$ beam-splitters (the success rate of the $1$-bit process when Bob inserts his detectors is above $95 \%$ if $N \geq 50$). Hence, Fig. \ref{fig:CMZIStrength} validifies the counterfactuality of the CMZI scheme for the values of $\theta_w$ that we have considered. We see that for small values of $\theta_w$ the value of $P_{\bm{A}}$ is kept well below the free-space interaction value of $1$. Regardless of the small values of $P_{\bm{A}}$, Fig.  \ref{fig:CMZIStrength} shows that Alice obtains a large amount of Fisher information $F_{\bm{A}}(\theta_w)$ about Bob's parameter $\theta_w$. The value of $F_{\bm{A}}(\theta)$ is independent of changes in the value of $\theta$ for small $\theta=\theta_w \approx 0$. The value of $P_{\bm{A}}$ is not.

\section{Concluding Remarks}

In this paper we have carried out a thorough study of the classical Fisher information and the  Shannon mutual information in nested Mach-Zehnder interferometers. We have calculated these information measures with respect to the measurement outcomes caused by polarization rotations in different parts of the structure. We find that in an otherwise non-polarizing optical circuit with real beam-splitter matrices, the Fisher information caused by a single polarization rotation is always proportional to the integrated probability density distribution at the location of the interaction in the Schr\"odinger picture. The Fisher information---in such a scenario---can thus be thought of as a measure of inter-measurement ``presence" at the rotator. Furthermore, we have developed \drma{interpretation independent} measures for the strength of counterfactual violations in two different types of CFC schemes. Type \rom{1} schemes do not allow particles to cross the transmission line between transmitter and receiver, whilst the Type \rom{2} schemes allow particles to travel in to opposite direction to the message. The rudimentary assumption made in this paper, is that any real quantum channel will naturally have unwanted components \drma{associated with} it. We introduce a small polarization rotation as a simple model of an unwanted quantum evolution in the devices. We find that a suitable measure of Type \rom{1} counterfactual violations is based on the classical Fisher information of the parameter $\theta_w$ that set the polarization rotation. It is scaled by the reciprocal Fisher information from a free-space interaction of the same strength as the one used in the device of interest. Hence, a value of $\mathcal{D}=1$ corresponds to the counterfactual violation strength of a free-space interaction inside the laboratory of study, and any value between $0$ and $1$ corresponds to a weaker-than-free-space counterfactual violation. For example, we can show that the suggested scheme of Salih et al. \cite{Salih13} strongly violates counterfactuality. Moreover, we provide an analytical study as well as a numerical simulation of the quantum evolution of a Gaussian particle in a single NMZI. The numerics show how the evolution of the quantum wavefunction leads to Fisher information of the interaction parameter (and ultimately the counterfactual violations) in NMZIs communication schemes. The analytical study supports the claim of the invalidity of counterfactual schemes based on NMZIs. The Type \rom{2} scheme developed by Arvidsson-Shukur and Barnes \cite{ArvShukur16} should be measured with another counterfactual violation measure. This measure, $P_{\bm{A}}$, is based on how much probability density that weak non-collapsing interactions in Bob's laboratory generate in Alice's. We find that the Type \rom{2} scheme of Ref. \cite{ArvShukur16} keeps the value of $P_{\bm{A}}$ to a fraction of a percent \drma{for weak interactions}, and thus satisfies its counterfactuality definition.


The authors would like to express their gratitude towards Thomas Bahder and Hugo Lepage for helpful discussions. This work was supported by the EPSRC, and the Cambridge Laboratory of Hitachi Limited via Project for Developing Innovation Systems of the MEXT in Japan. It has also received generous contributions from Helge Ax:son Johson’s Trust Fund, Anna Whitlock’s Trust Fund, Sixten Gemzeus’ Foundation, L\"angmanska kulturfonden’s Trust Fund, Tornspiran’s Trust Fund, Lars Hiertas’ Memorial Foundation and Gustaf S\"oderberg’s Trust Fund.

\bibliography{NestMZIInf}

\begin{thebibliography}{42}%
\makeatletter
\providecommand \@ifxundefined [1]{%
 \@ifx{#1\undefined}
}%
\providecommand \@ifnum [1]{%
 \ifnum #1\expandafter \@firstoftwo
 \else \expandafter \@secondoftwo
 \fi
}%
\providecommand \@ifx [1]{%
 \ifx #1\expandafter \@firstoftwo
 \else \expandafter \@secondoftwo
 \fi
}%
\providecommand \natexlab [1]{#1}%
\providecommand \enquote  [1]{``#1''}%
\providecommand \bibnamefont  [1]{#1}%
\providecommand \bibfnamefont [1]{#1}%
\providecommand \citenamefont [1]{#1}%
\providecommand \href@noop [0]{\@secondoftwo}%
\providecommand \href [0]{\begingroup \@sanitize@url \@href}%
\providecommand \@href[1]{\@@startlink{#1}\@@href}%
\providecommand \@@href[1]{\endgroup#1\@@endlink}%
\providecommand \@sanitize@url [0]{\catcode `\\12\catcode `\$12\catcode
  `\&12\catcode `\#12\catcode `\^12\catcode `\_12\catcode `\%12\relax}%
\providecommand \@@startlink[1]{}%
\providecommand \@@endlink[0]{}%
\providecommand \url  [0]{\begingroup\@sanitize@url \@url }%
\providecommand \@url [1]{\endgroup\@href {#1}{\urlprefix }}%
\providecommand \urlprefix  [0]{URL }%
\providecommand \Eprint [0]{\href }%
\providecommand \doibase [0]{http://dx.doi.org/}%
\providecommand \selectlanguage [0]{\@gobble}%
\providecommand \bibinfo  [0]{\@secondoftwo}%
\providecommand \bibfield  [0]{\@secondoftwo}%
\providecommand \translation [1]{[#1]}%
\providecommand \BibitemOpen [0]{}%
\providecommand \bibitemStop [0]{}%
\providecommand \bibitemNoStop [0]{.\EOS\space}%
\providecommand \EOS [0]{\spacefactor3000\relax}%
\providecommand \BibitemShut  [1]{\csname bibitem#1\endcsname}%
\let\auto@bib@innerbib\@empty
\bibitem [{\citenamefont {Salih}\ \emph {et~al.}(2013)\citenamefont {Salih},
  \citenamefont {Li}, \citenamefont {Al-Amri},\ and\ \citenamefont
  {Zubairy}}]{Salih13}%
  \BibitemOpen
  \bibfield  {author} {\bibinfo {author} {\bibfnamefont {H.}~\bibnamefont
  {Salih}}, \bibinfo {author} {\bibfnamefont {Z.-H.}\ \bibnamefont {Li}},
  \bibinfo {author} {\bibfnamefont {M.}~\bibnamefont {Al-Amri}}, \ and\
  \bibinfo {author} {\bibfnamefont {M.~S.}\ \bibnamefont {Zubairy}},\ }\href
  {\doibase 10.1103/PhysRevLett.110.170502} {\bibfield  {journal} {\bibinfo
  {journal} {Phys. Rev. Lett.}\ }\textbf {\bibinfo {volume} {110}},\ \bibinfo
  {pages} {170502} (\bibinfo {year} {2013})}\BibitemShut {NoStop}%
\bibitem [{\citenamefont {Arvidsson-Shukur}\ and\ \citenamefont
  {Barnes}(2016)}]{ArvShukur16}%
  \BibitemOpen
  \bibfield  {author} {\bibinfo {author} {\bibfnamefont {D.~R.~M.}\
  \bibnamefont {Arvidsson-Shukur}}\ and\ \bibinfo {author} {\bibfnamefont
  {C.~H.~W.}\ \bibnamefont {Barnes}},\ }\href {\doibase
  10.1103/PhysRevA.94.062303} {\bibfield  {journal} {\bibinfo  {journal} {Phys.
  Rev. A}\ }\textbf {\bibinfo {volume} {94}},\ \bibinfo {pages} {062303}
  (\bibinfo {year} {2016})}\BibitemShut {NoStop}%
\bibitem [{\citenamefont {Michelson}\ and\ \citenamefont
  {Morley}(1887)}]{Michelson1887}%
  \BibitemOpen
  \bibfield  {author} {\bibinfo {author} {\bibfnamefont {A.~A.}\ \bibnamefont
  {Michelson}}\ and\ \bibinfo {author} {\bibfnamefont {E.~W.}\ \bibnamefont
  {Morley}},\ }\href {\doibase 10.2475/ajs.s3-34.203.333} {\bibfield  {journal}
  {\bibinfo  {journal} {American Journal of Science}\ }\bibinfo {series} {3},\
  \textbf {\bibinfo {volume} {32}},\ \bibinfo {pages} {333} (\bibinfo {year}
  {1887})}\BibitemShut {NoStop}%
\bibitem [{\citenamefont {Hardy}(1992)}]{Hardy92}%
  \BibitemOpen
  \bibfield  {author} {\bibinfo {author} {\bibfnamefont {L.}~\bibnamefont
  {Hardy}},\ }\href {\doibase 10.1103/PhysRevLett.68.2981} {\bibfield
  {journal} {\bibinfo  {journal} {Phys. Rev. Lett.}\ }\textbf {\bibinfo
  {volume} {68}},\ \bibinfo {pages} {2981} (\bibinfo {year}
  {1992})}\BibitemShut {NoStop}%
\bibitem [{\citenamefont {Abbott}\ and\ \citenamefont
  {et~al.}(2016)}]{Abbott16}%
  \BibitemOpen
  \bibfield  {author} {\bibinfo {author} {\bibfnamefont {B.~P.}\ \bibnamefont
  {Abbott}}\ and\ \bibinfo {author} {\bibnamefont {et~al.}} (\bibinfo
  {collaboration} {LIGO Scientific Collaboration and Virgo Collaboration}),\
  }\href {\doibase 10.1103/PhysRevLett.116.061102} {\bibfield  {journal}
  {\bibinfo  {journal} {Phys. Rev. Lett.}\ }\textbf {\bibinfo {volume} {116}},\
  \bibinfo {pages} {061102} (\bibinfo {year} {2016})}\BibitemShut {NoStop}%
\bibitem [{\citenamefont {Hariharan}(2003)}]{Hariharan03}%
  \BibitemOpen
  \bibfield  {author} {\bibinfo {author} {\bibfnamefont {P.}~\bibnamefont
  {Hariharan}},\ }\href@noop {} {\emph {\bibinfo {title} {Optical
  Interferometry}}},\ \bibinfo {edition} {2nd}\ ed.\ (\bibinfo  {publisher}
  {Academic Press},\ \bibinfo {address} {Cambridge, Massachusetts, USA},\
  \bibinfo {year} {2003})\ p.\ \bibinfo {pages} {351}\BibitemShut {NoStop}%
\bibitem [{\citenamefont {Bertocchi}\ \emph {et~al.}(2006)\citenamefont
  {Bertocchi}, \citenamefont {Alibart}, \citenamefont {Ostrowsky},
  \citenamefont {Tanzilli},\ and\ \citenamefont {Baldi}}]{Bertocchi06}%
  \BibitemOpen
  \bibfield  {author} {\bibinfo {author} {\bibfnamefont {G.}~\bibnamefont
  {Bertocchi}}, \bibinfo {author} {\bibfnamefont {O.}~\bibnamefont {Alibart}},
  \bibinfo {author} {\bibfnamefont {D.~B.}\ \bibnamefont {Ostrowsky}}, \bibinfo
  {author} {\bibfnamefont {S.}~\bibnamefont {Tanzilli}}, \ and\ \bibinfo
  {author} {\bibfnamefont {P.}~\bibnamefont {Baldi}},\ }\href
  {http://stacks.iop.org/0953-4075/39/i=5/a=001} {\bibfield  {journal}
  {\bibinfo  {journal} {Journal of Physics B: Atomic, Molecular and Optical
  Physics}\ }\textbf {\bibinfo {volume} {39}},\ \bibinfo {pages} {1011}
  (\bibinfo {year} {2006})}\BibitemShut {NoStop}%
\bibitem [{\citenamefont {Berti}\ \emph {et~al.}(2007)\citenamefont {Berti},
  \citenamefont {Cardoso}, \citenamefont {Cardoso},\ and\ \citenamefont
  {Cavagli\`a}}]{Berti07}%
  \BibitemOpen
  \bibfield  {author} {\bibinfo {author} {\bibfnamefont {E.}~\bibnamefont
  {Berti}}, \bibinfo {author} {\bibfnamefont {J.}~\bibnamefont {Cardoso}},
  \bibinfo {author} {\bibfnamefont {V.}~\bibnamefont {Cardoso}}, \ and\
  \bibinfo {author} {\bibfnamefont {M.}~\bibnamefont {Cavagli\`a}},\ }\href
  {\doibase 10.1103/PhysRevD.76.104044} {\bibfield  {journal} {\bibinfo
  {journal} {Phys. Rev. D}\ }\textbf {\bibinfo {volume} {76}},\ \bibinfo
  {pages} {104044} (\bibinfo {year} {2007})}\BibitemShut {NoStop}%
\bibitem [{\citenamefont {Stockton}\ \emph {et~al.}(2007)\citenamefont
  {Stockton}, \citenamefont {Wu},\ and\ \citenamefont {Kasevich}}]{Stockton07}%
  \BibitemOpen
  \bibfield  {author} {\bibinfo {author} {\bibfnamefont {J.~K.}\ \bibnamefont
  {Stockton}}, \bibinfo {author} {\bibfnamefont {X.}~\bibnamefont {Wu}}, \ and\
  \bibinfo {author} {\bibfnamefont {M.~A.}\ \bibnamefont {Kasevich}},\ }\href
  {\doibase 10.1103/PhysRevA.76.033613} {\bibfield  {journal} {\bibinfo
  {journal} {Phys. Rev. A}\ }\textbf {\bibinfo {volume} {76}},\ \bibinfo
  {pages} {033613} (\bibinfo {year} {2007})}\BibitemShut {NoStop}%
\bibitem [{\citenamefont {Demkowicz-Dobrzanski}\ \emph
  {et~al.}(2009)\citenamefont {Demkowicz-Dobrzanski}, \citenamefont {Dorner},
  \citenamefont {Smith}, \citenamefont {Lundeen}, \citenamefont {Wasilewski},
  \citenamefont {Banaszek},\ and\ \citenamefont {Walmsley}}]{Dorner09}%
  \BibitemOpen
  \bibfield  {author} {\bibinfo {author} {\bibfnamefont {R.}~\bibnamefont
  {Demkowicz-Dobrzanski}}, \bibinfo {author} {\bibfnamefont {U.}~\bibnamefont
  {Dorner}}, \bibinfo {author} {\bibfnamefont {B.~J.}\ \bibnamefont {Smith}},
  \bibinfo {author} {\bibfnamefont {J.~S.}\ \bibnamefont {Lundeen}}, \bibinfo
  {author} {\bibfnamefont {W.}~\bibnamefont {Wasilewski}}, \bibinfo {author}
  {\bibfnamefont {K.}~\bibnamefont {Banaszek}}, \ and\ \bibinfo {author}
  {\bibfnamefont {I.~A.}\ \bibnamefont {Walmsley}},\ }\href {\doibase
  10.1103/PhysRevA.80.013825} {\bibfield  {journal} {\bibinfo  {journal} {Phys.
  Rev. A}\ }\textbf {\bibinfo {volume} {80}},\ \bibinfo {pages} {013825}
  (\bibinfo {year} {2009})}\BibitemShut {NoStop}%
\bibitem [{\citenamefont {Cable}\ and\ \citenamefont {Durkin}(2010)}]{Cable10}%
  \BibitemOpen
  \bibfield  {author} {\bibinfo {author} {\bibfnamefont {H.}~\bibnamefont
  {Cable}}\ and\ \bibinfo {author} {\bibfnamefont {G.~A.}\ \bibnamefont
  {Durkin}},\ }\href {\doibase 10.1103/PhysRevLett.105.013603} {\bibfield
  {journal} {\bibinfo  {journal} {Phys. Rev. Lett.}\ }\textbf {\bibinfo
  {volume} {105}},\ \bibinfo {pages} {013603} (\bibinfo {year}
  {2010})}\BibitemShut {NoStop}%
\bibitem [{\citenamefont {Vitale}\ and\ \citenamefont
  {Zanolin}(2010)}]{Vitale10}%
  \BibitemOpen
  \bibfield  {author} {\bibinfo {author} {\bibfnamefont {S.}~\bibnamefont
  {Vitale}}\ and\ \bibinfo {author} {\bibfnamefont {M.}~\bibnamefont
  {Zanolin}},\ }\href {\doibase 10.1103/PhysRevD.82.124065} {\bibfield
  {journal} {\bibinfo  {journal} {Phys. Rev. D}\ }\textbf {\bibinfo {volume}
  {82}},\ \bibinfo {pages} {124065} (\bibinfo {year} {2010})}\BibitemShut
  {NoStop}%
\bibitem [{\citenamefont {Bahder}(2011)}]{Bahder11}%
  \BibitemOpen
  \bibfield  {author} {\bibinfo {author} {\bibfnamefont {T.~B.}\ \bibnamefont
  {Bahder}},\ }\href {\doibase 10.1103/PhysRevA.83.053601} {\bibfield
  {journal} {\bibinfo  {journal} {Phys. Rev. A}\ }\textbf {\bibinfo {volume}
  {83}},\ \bibinfo {pages} {053601} (\bibinfo {year} {2011})}\BibitemShut
  {NoStop}%
\bibitem [{\citenamefont {Rozema}\ \emph
  {et~al.}(2012{\natexlab{a}})\citenamefont {Rozema}, \citenamefont {Darabi},
  \citenamefont {Mahler}, \citenamefont {Hayat}, \citenamefont {Soudagar},\
  and\ \citenamefont {Steinberg}}]{Rozema12}%
  \BibitemOpen
  \bibfield  {author} {\bibinfo {author} {\bibfnamefont {L.}~\bibnamefont
  {Rozema}}, \bibinfo {author} {\bibfnamefont {A.}~\bibnamefont {Darabi}},
  \bibinfo {author} {\bibfnamefont {D.}~\bibnamefont {Mahler}}, \bibinfo
  {author} {\bibfnamefont {A.}~\bibnamefont {Hayat}}, \bibinfo {author}
  {\bibfnamefont {Y.}~\bibnamefont {Soudagar}}, \ and\ \bibinfo {author}
  {\bibfnamefont {A.~M.}\ \bibnamefont {Steinberg}},\ }in\ \href {\doibase
  10.1364/FIO.2012.FW4J.4} {\emph {\bibinfo {booktitle} {Frontiers in Optics
  2012/Laser Science XXVIII}}}\ (\bibinfo  {publisher} {Optical Society of
  America},\ \bibinfo {year} {2012})\ p.\ \bibinfo {pages} {FW4J.4}\BibitemShut
  {NoStop}%
\bibitem [{\citenamefont {Rozema}\ \emph
  {et~al.}(2012{\natexlab{b}})\citenamefont {Rozema}, \citenamefont {Darabi},
  \citenamefont {Mahler}, \citenamefont {Hayat}, \citenamefont {Soudagar},\
  and\ \citenamefont {Steinberg}}]{Rozema12-2}%
  \BibitemOpen
  \bibfield  {author} {\bibinfo {author} {\bibfnamefont {L.~A.}\ \bibnamefont
  {Rozema}}, \bibinfo {author} {\bibfnamefont {A.}~\bibnamefont {Darabi}},
  \bibinfo {author} {\bibfnamefont {D.~H.}\ \bibnamefont {Mahler}}, \bibinfo
  {author} {\bibfnamefont {A.}~\bibnamefont {Hayat}}, \bibinfo {author}
  {\bibfnamefont {Y.}~\bibnamefont {Soudagar}}, \ and\ \bibinfo {author}
  {\bibfnamefont {A.~M.}\ \bibnamefont {Steinberg}},\ }\href {\doibase
  10.1103/PhysRevLett.109.100404} {\bibfield  {journal} {\bibinfo  {journal}
  {Phys. Rev. Lett.}\ }\textbf {\bibinfo {volume} {109}},\ \bibinfo {pages}
  {100404} (\bibinfo {year} {2012}{\natexlab{b}})}\BibitemShut {NoStop}%
\bibitem [{\citenamefont {Mahler}\ \emph {et~al.}(2013)\citenamefont {Mahler},
  \citenamefont {Rozema}, \citenamefont {Darabi}, \citenamefont {Ferrie},
  \citenamefont {Blume-Kohout},\ and\ \citenamefont {Steinberg}}]{Mahler13}%
  \BibitemOpen
  \bibfield  {author} {\bibinfo {author} {\bibfnamefont {D.~H.}\ \bibnamefont
  {Mahler}}, \bibinfo {author} {\bibfnamefont {L.~A.}\ \bibnamefont {Rozema}},
  \bibinfo {author} {\bibfnamefont {A.}~\bibnamefont {Darabi}}, \bibinfo
  {author} {\bibfnamefont {C.}~\bibnamefont {Ferrie}}, \bibinfo {author}
  {\bibfnamefont {R.}~\bibnamefont {Blume-Kohout}}, \ and\ \bibinfo {author}
  {\bibfnamefont {A.~M.}\ \bibnamefont {Steinberg}},\ }\href {\doibase
  10.1103/PhysRevLett.111.183601} {\bibfield  {journal} {\bibinfo  {journal}
  {Phys. Rev. Lett.}\ }\textbf {\bibinfo {volume} {111}},\ \bibinfo {pages}
  {183601} (\bibinfo {year} {2013})}\BibitemShut {NoStop}%
\bibitem [{\citenamefont {Hradil}(1995)}]{Hradil95}%
  \BibitemOpen
  \bibfield  {author} {\bibinfo {author} {\bibfnamefont {Z.}~\bibnamefont
  {Hradil}},\ }\href {\doibase 10.1103/PhysRevA.51.1870} {\bibfield  {journal}
  {\bibinfo  {journal} {Phys. Rev. A}\ }\textbf {\bibinfo {volume} {51}},\
  \bibinfo {pages} {1870} (\bibinfo {year} {1995})}\BibitemShut {NoStop}%
\bibitem [{\citenamefont {Bahder}\ and\ \citenamefont
  {Lopata}(2006)}]{Bahder06}%
  \BibitemOpen
  \bibfield  {author} {\bibinfo {author} {\bibfnamefont {T.~B.}\ \bibnamefont
  {Bahder}}\ and\ \bibinfo {author} {\bibfnamefont {P.~A.}\ \bibnamefont
  {Lopata}},\ }\href {\doibase 10.1103/PhysRevA.74.051801} {\bibfield
  {journal} {\bibinfo  {journal} {Phys. Rev. A}\ }\textbf {\bibinfo {volume}
  {74}},\ \bibinfo {pages} {051801} (\bibinfo {year} {2006})}\BibitemShut
  {NoStop}%
\bibitem [{\citenamefont {Simon}\ \emph {et~al.}(2008)\citenamefont {Simon},
  \citenamefont {Sergienko},\ and\ \citenamefont {Bahder}}]{Simon08}%
  \BibitemOpen
  \bibfield  {author} {\bibinfo {author} {\bibfnamefont {D.~S.}\ \bibnamefont
  {Simon}}, \bibinfo {author} {\bibfnamefont {A.~V.}\ \bibnamefont
  {Sergienko}}, \ and\ \bibinfo {author} {\bibfnamefont {T.~B.}\ \bibnamefont
  {Bahder}},\ }\href {\doibase 10.1103/PhysRevA.78.053829} {\bibfield
  {journal} {\bibinfo  {journal} {Phys. Rev. A}\ }\textbf {\bibinfo {volume}
  {78}},\ \bibinfo {pages} {053829} (\bibinfo {year} {2008})}\BibitemShut
  {NoStop}%
\bibitem [{\citenamefont {Penrose}(1994)}]{bPenrose94}%
  \BibitemOpen
  \bibfield  {author} {\bibinfo {author} {\bibfnamefont {R.}~\bibnamefont
  {Penrose}},\ }\href@noop {} {\emph {\bibinfo {title} {Shadows of the Mind: A
  Search for the Missing Science of Consciousness}}},\ \bibinfo {edition}
  {1st}\ ed.\ (\bibinfo  {publisher} {Oxford University Press, Inc.},\ \bibinfo
  {address} {New York, NY, USA},\ \bibinfo {year} {1994})\ p.\ \bibinfo {pages}
  {240}\BibitemShut {NoStop}%
\bibitem [{\citenamefont {Elitzur}\ and\ \citenamefont
  {Vaidman}(1993)}]{Elitzur93}%
  \BibitemOpen
  \bibfield  {author} {\bibinfo {author} {\bibfnamefont {A.~C.}\ \bibnamefont
  {Elitzur}}\ and\ \bibinfo {author} {\bibfnamefont {L.}~\bibnamefont
  {Vaidman}},\ }\href {\doibase 10.1007/BF00736012} {\bibfield  {journal}
  {\bibinfo  {journal} {Foundations of Physics}\ }\textbf {\bibinfo {volume}
  {23}},\ \bibinfo {pages} {987} (\bibinfo {year} {1993})}\BibitemShut
  {NoStop}%
\bibitem [{\citenamefont {Kwiat}\ \emph {et~al.}(1995)\citenamefont {Kwiat},
  \citenamefont {Weinfurter}, \citenamefont {Herzog}, \citenamefont
  {Zeilinger},\ and\ \citenamefont {Kasevich}}]{Kwiat95}%
  \BibitemOpen
  \bibfield  {author} {\bibinfo {author} {\bibfnamefont {P.}~\bibnamefont
  {Kwiat}}, \bibinfo {author} {\bibfnamefont {H.}~\bibnamefont {Weinfurter}},
  \bibinfo {author} {\bibfnamefont {T.}~\bibnamefont {Herzog}}, \bibinfo
  {author} {\bibfnamefont {A.}~\bibnamefont {Zeilinger}}, \ and\ \bibinfo
  {author} {\bibfnamefont {M.~A.}\ \bibnamefont {Kasevich}},\ }\href {\doibase
  10.1103/PhysRevLett.74.4763} {\bibfield  {journal} {\bibinfo  {journal}
  {Phys. Rev. Lett.}\ }\textbf {\bibinfo {volume} {74}},\ \bibinfo {pages}
  {4763} (\bibinfo {year} {1995})}\BibitemShut {NoStop}%
\bibitem [{\citenamefont {Kwiat}\ \emph {et~al.}(1999)\citenamefont {Kwiat},
  \citenamefont {White}, \citenamefont {Mitchell}, \citenamefont {Nairz},
  \citenamefont {Weihs}, \citenamefont {Weinfurter},\ and\ \citenamefont
  {Zeilinger}}]{Kwiat99}%
  \BibitemOpen
  \bibfield  {author} {\bibinfo {author} {\bibfnamefont {P.~G.}\ \bibnamefont
  {Kwiat}}, \bibinfo {author} {\bibfnamefont {A.~G.}\ \bibnamefont {White}},
  \bibinfo {author} {\bibfnamefont {J.~R.}\ \bibnamefont {Mitchell}}, \bibinfo
  {author} {\bibfnamefont {O.}~\bibnamefont {Nairz}}, \bibinfo {author}
  {\bibfnamefont {G.}~\bibnamefont {Weihs}}, \bibinfo {author} {\bibfnamefont
  {H.}~\bibnamefont {Weinfurter}}, \ and\ \bibinfo {author} {\bibfnamefont
  {A.}~\bibnamefont {Zeilinger}},\ }\href {\doibase
  10.1103/PhysRevLett.83.4725} {\bibfield  {journal} {\bibinfo  {journal}
  {Phys. Rev. Lett.}\ }\textbf {\bibinfo {volume} {83}},\ \bibinfo {pages}
  {4725} (\bibinfo {year} {1999})}\BibitemShut {NoStop}%
\bibitem [{\citenamefont {Hosten}\ \emph {et~al.}(2006)\citenamefont {Hosten},
  \citenamefont {Rakher}, \citenamefont {Barreiro}, \citenamefont {Peters},\
  and\ \citenamefont {Kwiat}}]{Hosten06}%
  \BibitemOpen
  \bibfield  {author} {\bibinfo {author} {\bibfnamefont {O.}~\bibnamefont
  {Hosten}}, \bibinfo {author} {\bibfnamefont {M.~T.}\ \bibnamefont {Rakher}},
  \bibinfo {author} {\bibfnamefont {J.~T.}\ \bibnamefont {Barreiro}}, \bibinfo
  {author} {\bibfnamefont {N.~A.}\ \bibnamefont {Peters}}, \ and\ \bibinfo
  {author} {\bibfnamefont {P.~G.}\ \bibnamefont {Kwiat}},\ }\href {\doibase
  http://dx.doi.org/10.1038/nature04523} {\bibfield  {journal} {\bibinfo
  {journal} {Nature}\ }\textbf {\bibinfo {volume} {439}} (\bibinfo {year}
  {2006}),\ http://dx.doi.org/10.1038/nature04523}\BibitemShut {NoStop}%
\bibitem [{\citenamefont {Cao}\ \emph {et~al.}(2017)\citenamefont {Cao},
  \citenamefont {Li}, \citenamefont {Cao}, \citenamefont {Yin}, \citenamefont
  {Chen}, \citenamefont {Yin}, \citenamefont {Chen}, \citenamefont {Ma},
  \citenamefont {Peng},\ and\ \citenamefont {Pan}}]{Cao17}%
  \BibitemOpen
  \bibfield  {author} {\bibinfo {author} {\bibfnamefont {Y.}~\bibnamefont
  {Cao}}, \bibinfo {author} {\bibfnamefont {Y.-H.}\ \bibnamefont {Li}},
  \bibinfo {author} {\bibfnamefont {Z.}~\bibnamefont {Cao}}, \bibinfo {author}
  {\bibfnamefont {J.}~\bibnamefont {Yin}}, \bibinfo {author} {\bibfnamefont
  {Y.-A.}\ \bibnamefont {Chen}}, \bibinfo {author} {\bibfnamefont {H.-L.}\
  \bibnamefont {Yin}}, \bibinfo {author} {\bibfnamefont {T.-Y.}\ \bibnamefont
  {Chen}}, \bibinfo {author} {\bibfnamefont {X.}~\bibnamefont {Ma}}, \bibinfo
  {author} {\bibfnamefont {C.-Z.}\ \bibnamefont {Peng}}, \ and\ \bibinfo
  {author} {\bibfnamefont {J.-W.}\ \bibnamefont {Pan}},\ }\href@noop {}
  {\bibfield  {journal} {\bibinfo  {journal} {Proceedings of the National
  Academy of Sciences}\ }\textbf {\bibinfo {volume} {114}},\ \bibinfo {pages}
  {4920} (\bibinfo {year} {2017})}\BibitemShut {NoStop}%
\bibitem [{\citenamefont {Li}\ \emph {et~al.}(2015)\citenamefont {Li},
  \citenamefont {Al-Amri},\ and\ \citenamefont {Zubairy}}]{Li15}%
  \BibitemOpen
  \bibfield  {author} {\bibinfo {author} {\bibfnamefont {Z.-H.}\ \bibnamefont
  {Li}}, \bibinfo {author} {\bibfnamefont {M.}~\bibnamefont {Al-Amri}}, \ and\
  \bibinfo {author} {\bibfnamefont {M.~S.}\ \bibnamefont {Zubairy}},\ }\href
  {\doibase 10.1103/PhysRevA.92.052315} {\bibfield  {journal} {\bibinfo
  {journal} {Phys. Rev. A}\ }\textbf {\bibinfo {volume} {92}},\ \bibinfo
  {pages} {052315} (\bibinfo {year} {2015})}\BibitemShut {NoStop}%
\bibitem [{\citenamefont {Vaidman}(2013{\natexlab{a}})}]{Vaidman13}%
  \BibitemOpen
  \bibfield  {author} {\bibinfo {author} {\bibfnamefont {L.}~\bibnamefont
  {Vaidman}},\ }\href {\doibase 10.1103/PhysRevA.87.052104} {\bibfield
  {journal} {\bibinfo  {journal} {Phys. Rev. A}\ }\textbf {\bibinfo {volume}
  {87}},\ \bibinfo {pages} {052104} (\bibinfo {year}
  {2013}{\natexlab{a}})}\BibitemShut {NoStop}%
\bibitem [{\citenamefont {Li}\ \emph {et~al.}(2013)\citenamefont {Li},
  \citenamefont {Al-Amri},\ and\ \citenamefont {Zubairy}}]{Li13}%
  \BibitemOpen
  \bibfield  {author} {\bibinfo {author} {\bibfnamefont {Z.-H.}\ \bibnamefont
  {Li}}, \bibinfo {author} {\bibfnamefont {M.}~\bibnamefont {Al-Amri}}, \ and\
  \bibinfo {author} {\bibfnamefont {M.~S.}\ \bibnamefont {Zubairy}},\ }\href
  {\doibase 10.1103/PhysRevA.88.046102} {\bibfield  {journal} {\bibinfo
  {journal} {Phys. Rev. A}\ }\textbf {\bibinfo {volume} {88}},\ \bibinfo
  {pages} {046102} (\bibinfo {year} {2013})}\BibitemShut {NoStop}%
\bibitem [{\citenamefont {Vaidman}(2013{\natexlab{b}})}]{Vaidman13-2}%
  \BibitemOpen
  \bibfield  {author} {\bibinfo {author} {\bibfnamefont {L.}~\bibnamefont
  {Vaidman}},\ }\href {\doibase 10.1103/PhysRevA.88.046103} {\bibfield
  {journal} {\bibinfo  {journal} {Phys. Rev. A}\ }\textbf {\bibinfo {volume}
  {88}},\ \bibinfo {pages} {046103} (\bibinfo {year}
  {2013}{\natexlab{b}})}\BibitemShut {NoStop}%
\bibitem [{\citenamefont {Vaidman}(2014)}]{Vaidman14}%
  \BibitemOpen
  \bibfield  {author} {\bibinfo {author} {\bibfnamefont {L.}~\bibnamefont
  {Vaidman}},\ }\href {\doibase 10.1103/PhysRevLett.112.208901} {\bibfield
  {journal} {\bibinfo  {journal} {Phys. Rev. Lett.}\ }\textbf {\bibinfo
  {volume} {112}},\ \bibinfo {pages} {208901} (\bibinfo {year}
  {2014})}\BibitemShut {NoStop}%
\bibitem [{\citenamefont {Salih}\ \emph {et~al.}(2014)\citenamefont {Salih},
  \citenamefont {Li}, \citenamefont {Al-Amri},\ and\ \citenamefont
  {Zubairy}}]{Salih14}%
  \BibitemOpen
  \bibfield  {author} {\bibinfo {author} {\bibfnamefont {H.}~\bibnamefont
  {Salih}}, \bibinfo {author} {\bibfnamefont {Z.-H.}\ \bibnamefont {Li}},
  \bibinfo {author} {\bibfnamefont {M.}~\bibnamefont {Al-Amri}}, \ and\
  \bibinfo {author} {\bibfnamefont {M.~S.}\ \bibnamefont {Zubairy}},\ }\href
  {\doibase 10.1103/PhysRevLett.112.208902} {\bibfield  {journal} {\bibinfo
  {journal} {Phys. Rev. Lett.}\ }\textbf {\bibinfo {volume} {112}},\ \bibinfo
  {pages} {208902} (\bibinfo {year} {2014})}\BibitemShut {NoStop}%
\bibitem [{\citenamefont {Vaidman}(2016)}]{Vaidman16}%
  \BibitemOpen
  \bibfield  {author} {\bibinfo {author} {\bibfnamefont {L.}~\bibnamefont
  {Vaidman}},\ }\href {\doibase 10.1103/PhysRevA.93.066301} {\bibfield
  {journal} {\bibinfo  {journal} {Phys. Rev. A}\ }\textbf {\bibinfo {volume}
  {93}},\ \bibinfo {pages} {066301} (\bibinfo {year} {2016})}\BibitemShut
  {NoStop}%
\bibitem [{\citenamefont {Li}\ \emph {et~al.}(2016)\citenamefont {Li},
  \citenamefont {Al-Amri},\ and\ \citenamefont {Zubairy}}]{Li16}%
  \BibitemOpen
  \bibfield  {author} {\bibinfo {author} {\bibfnamefont {Z.-H.}\ \bibnamefont
  {Li}}, \bibinfo {author} {\bibfnamefont {M.}~\bibnamefont {Al-Amri}}, \ and\
  \bibinfo {author} {\bibfnamefont {M.~S.}\ \bibnamefont {Zubairy}},\ }\href
  {\doibase 10.1103/PhysRevA.93.066302} {\bibfield  {journal} {\bibinfo
  {journal} {Phys. Rev. A}\ }\textbf {\bibinfo {volume} {93}},\ \bibinfo
  {pages} {066302} (\bibinfo {year} {2016})}\BibitemShut {NoStop}%
\bibitem [{\citenamefont {Griffiths}(2016)}]{Griffiths16}%
  \BibitemOpen
  \bibfield  {author} {\bibinfo {author} {\bibfnamefont {R.~B.}\ \bibnamefont
  {Griffiths}},\ }\href {\doibase 10.1103/PhysRevA.94.032115} {\bibfield
  {journal} {\bibinfo  {journal} {Phys. Rev. A}\ }\textbf {\bibinfo {volume}
  {94}},\ \bibinfo {pages} {032115} (\bibinfo {year} {2016})}\BibitemShut
  {NoStop}%
\bibitem [{\citenamefont {Cover}\ and\ \citenamefont
  {Thomas}(2006)}]{bCover06}%
  \BibitemOpen
  \bibfield  {author} {\bibinfo {author} {\bibfnamefont {T.~M.}\ \bibnamefont
  {Cover}}\ and\ \bibinfo {author} {\bibfnamefont {J.~A.}\ \bibnamefont
  {Thomas}},\ }\href@noop {} {\emph {\bibinfo {title} {Elements of Information
  Theory}}},\ \bibinfo {edition} {2nd}\ ed.\ (\bibinfo  {publisher} {John Wiley
  and Sons Inc.},\ \bibinfo {address} {Hoboken, New Jersey, USA},\ \bibinfo
  {year} {2006})\BibitemShut {NoStop}%
\bibitem [{\citenamefont {Marlow}(1978)}]{bMarlow78}%
  \BibitemOpen
  \bibfield  {author} {\bibinfo {author} {\bibfnamefont {A.}~\bibnamefont
  {Marlow}},\ }\href@noop {} {\emph {\bibinfo {title} {Mathematical Foundations
  of Quantum Theory}}},\ \bibinfo {edition} {1st}\ ed.\ (\bibinfo  {publisher}
  {ACADEMIC PRESS, INC. (LONDON) LTD.},\ \bibinfo {address} {London, UK},\
  \bibinfo {year} {1978})\BibitemShut {NoStop}%
\bibitem [{\citenamefont {Degasperis}\ \emph {et~al.}(1974)\citenamefont
  {Degasperis}, \citenamefont {Fonda},\ and\ \citenamefont
  {Ghirardi}}]{Degasperis74}%
  \BibitemOpen
  \bibfield  {author} {\bibinfo {author} {\bibfnamefont {A.}~\bibnamefont
  {Degasperis}}, \bibinfo {author} {\bibfnamefont {L.}~\bibnamefont {Fonda}}, \
  and\ \bibinfo {author} {\bibfnamefont {G.~C.}\ \bibnamefont {Ghirardi}},\
  }\href {\doibase 10.1007/BF02731351} {\bibfield  {journal} {\bibinfo
  {journal} {Il Nuovo Cimento A (1965-1970)}\ }\textbf {\bibinfo {volume}
  {21}},\ \bibinfo {pages} {471} (\bibinfo {year} {1974})}\BibitemShut
  {NoStop}%
\bibitem [{\citenamefont {Misra}\ and\ \citenamefont
  {Sudarshan}(1977)}]{Misra77}%
  \BibitemOpen
  \bibfield  {author} {\bibinfo {author} {\bibfnamefont {B.}~\bibnamefont
  {Misra}}\ and\ \bibinfo {author} {\bibfnamefont {E.~C.~G.}\ \bibnamefont
  {Sudarshan}},\ }\href@noop {} {\bibfield  {journal} {\bibinfo  {journal}
  {Journal of Mathematical Physics}\ }\textbf {\bibinfo {volume} {18}}
  (\bibinfo {year} {1977})}\BibitemShut {NoStop}%
\bibitem [{\citenamefont {Press}\ \emph {et~al.}(1992)\citenamefont {Press},
  \citenamefont {Teukolsky}, \citenamefont {Vetterling},\ and\ \citenamefont
  {Flannery}}]{bPress92}%
  \BibitemOpen
  \bibfield  {author} {\bibinfo {author} {\bibfnamefont {W.~H.}\ \bibnamefont
  {Press}}, \bibinfo {author} {\bibfnamefont {S.~A.}\ \bibnamefont
  {Teukolsky}}, \bibinfo {author} {\bibfnamefont {W.~T.}\ \bibnamefont
  {Vetterling}}, \ and\ \bibinfo {author} {\bibfnamefont {B.~P.}\ \bibnamefont
  {Flannery}},\ }\href@noop {} {\emph {\bibinfo {title} {Numerical Recipes in
  Fortran}}},\ \bibinfo {edition} {2nd}\ ed.,\ Vol.~\bibinfo {volume} {1}\
  (\bibinfo  {publisher} {Press Syndicate of the University of Cambridge},\
  \bibinfo {address} {Cambridge, UK},\ \bibinfo {year} {1992})\BibitemShut
  {NoStop}%
\bibitem [{\citenamefont {Vaidman}(2015)}]{Vaidman15}%
  \BibitemOpen
  \bibfield  {author} {\bibinfo {author} {\bibfnamefont {L.}~\bibnamefont
  {Vaidman}},\ }\href {http://stacks.iop.org/1751-8121/48/i=46/a=465303}
  {\bibfield  {journal} {\bibinfo  {journal} {Journal of Physics A:
  Mathematical and Theoretical}\ }\textbf {\bibinfo {volume} {48}},\ \bibinfo
  {pages} {465303} (\bibinfo {year} {2015})}\BibitemShut {NoStop}%
\bibitem [{\citenamefont {Owen}\ \emph {et~al.}(2014)\citenamefont {Owen},
  \citenamefont {Dean},\ and\ \citenamefont {Barnes}}]{Owen14}%
  \BibitemOpen
  \bibfield  {author} {\bibinfo {author} {\bibfnamefont {E.~T.}\ \bibnamefont
  {Owen}}, \bibinfo {author} {\bibfnamefont {M.~C.}\ \bibnamefont {Dean}}, \
  and\ \bibinfo {author} {\bibfnamefont {C.~H.~W.}\ \bibnamefont {Barnes}},\
  }\href {\doibase 10.1103/PhysRevA.89.032305} {\bibfield  {journal} {\bibinfo
  {journal} {Phys. Rev. A}\ }\textbf {\bibinfo {volume} {89}},\ \bibinfo
  {pages} {032305} (\bibinfo {year} {2014})}\BibitemShut {NoStop}%
\bibitem [{\citenamefont {Arvidsson-Shukur}\ \emph {et~al.}(2017)\citenamefont
  {Arvidsson-Shukur}, \citenamefont {Lepage}, \citenamefont {Owen},
  \citenamefont {Ferrus},\ and\ \citenamefont {Barnes}}]{ArvShukur17}%
  \BibitemOpen
  \bibfield  {author} {\bibinfo {author} {\bibfnamefont {D.~R.~M.}\
  \bibnamefont {Arvidsson-Shukur}}, \bibinfo {author} {\bibfnamefont {H.~V.}\
  \bibnamefont {Lepage}}, \bibinfo {author} {\bibfnamefont {E.~T.}\
  \bibnamefont {Owen}}, \bibinfo {author} {\bibfnamefont {T.}~\bibnamefont
  {Ferrus}}, \ and\ \bibinfo {author} {\bibfnamefont {C.~H.~W.}\ \bibnamefont
  {Barnes}},\ }\href {\doibase 10.1103/PhysRevA.96.052305} {\bibfield
  {journal} {\bibinfo  {journal} {Phys. Rev. A}\ }\textbf {\bibinfo {volume}
  {96}},\ \bibinfo {pages} {052305} (\bibinfo {year} {2017})}\BibitemShut
  {NoStop}%
\end{thebibliography}%

\end{document}